\documentclass[twocolumn]{aastex63}
\usepackage[utf8]{inputenc}
\usepackage{CJKutf8}
\usepackage{amsmath}
\submitjournal{The Astrophysical Journal, Jun 10 2022, Accepted Nov 20 2022}
\shorttitle{Analysis of T\.ZO \& sAGB Candidates}
\shortauthors{O`Grady et al.}

\begin{document}
\begin{CJK*}{UTF8}{gbsn}

\title{Cool, Luminous, and Highly Variable Stars in the Magellanic Clouds. II: Spectroscopic and Environmental Analysis of Thorne-\.Zytkow Object and Super-AGB Star Candidates}
\correspondingauthor{Anna O`Grady}
\email{anna.ogrady@mail.utoronto.ca}

\author[0000-0002-7296-6547]{Anna J. G. O`Grady}
\affil{David A. Dunlap Department of Astronomy \& Astrophysics, University of Toronto, 50 St. George Street, Toronto, ON, M5S3H4 Canada}
\affil{Dunlap Institute for Astronomy \& Astrophysics, University of Toronto, 50 St. George Street, Toronto, ON, M5S3H4, Canada}

\author[0000-0001-7081-0082]{Maria R. Drout}
\affil{David A. Dunlap Department of Astronomy \& Astrophysics, University of Toronto, 50 St. George Street, Toronto, ON, M5S3H4 Canada}
\affiliation{The Observatories of the Carnegie Institution for Science, 813 Santa Barbara St., Pasadena, CA 91101, USA}

\author[0000-0002-3382-9558]{B. M. Gaensler}
\affil{David A. Dunlap Department of Astronomy \& Astrophysics, University of Toronto, 50 St. George Street, Toronto, ON, M5S3H4 Canada}
\affil{Dunlap Institute for Astronomy \& Astrophysics, University of Toronto, 50 St. George Street, Toronto, ON, M5S3H4, Canada}

\author{C. S. Kochanek}
\affil{Department of Astronomy, The Ohio State University, 140 West 18th Avenue, Columbus, OH 43210, USA}
\affil{Center for Cosmology and Astroparticle Physics, The Ohio State University, 191 W. Woodruff Avenue, Columbus, OH 43210, USA}

\author[0000-0002-5787-138X]{Kathryn F. Neugent}
\altaffiliation{Hubble Fellow}
\affil{Center for Astrophysics | Harvard \& Smithsonian, 60 Garden Street, Cambridge, MA 02138-1516, USA}
\affil{Lowell Observatory, 1400 W Mars Hill Road, Flagstaff, AZ 86001, USA}
\affil{Dunlap Institute for Astronomy \& Astrophysics, University of Toronto, 50 St. George Street, Toronto, ON, M5S3H4, Canada}

\author[0000-0001-8161-4087]{Carolyn L. Doherty}
\affil{School of Physics and Astronomy, Monash University, VIC 3800, Australia}

\author[0000-0003-2573-9832]{Joshua S. Speagle (沈佳士)}
\affil{David A. Dunlap Department of Astronomy \& Astrophysics, University of Toronto, 50 St. George Street, Toronto, ON, M5S3H4 Canada}
\affiliation{Department of Statistical Sciences, University of Toronto, 100 St George Street, Toronto ON, M5S3G3, Canada}
\affil{Dunlap Institute for Astronomy \& Astrophysics, University of Toronto, 50 St. George Street, Toronto, ON, M5S3H4, Canada}

\author[0000-0003-4631-1149]{B. J. Shappee}
\affil{Institute for Astronomy, University of Hawaii, 2680 Woodlawn Drive, Honolulu, HI 96822,USA}

\author[0000-0002-1690-3488]{Michael Rauch}
\affiliation{The Observatories of the Carnegie Institution for Science, 813 Santa Barbara St., Pasadena, CA 91101, USA}

\author[0000-0002-6960-6911]{Ylva G\"{o}tberg}
\altaffiliation{Hubble Fellow}
\affil{The Observatories of the Carnegie Institution for Science, 813 Santa Barbara St., Pasadena, CA 91101, USA}

\author[0000-0003-0857-2989]{Bethany Ludwig}
\affil{David A. Dunlap Department of Astronomy \& Astrophysics, University of Toronto, 50 St. George Street, Toronto, ON, M5S3H4 Canada}
\affil{Dunlap Institute for Astronomy \& Astrophysics, University of Toronto, 50 St. George Street, Toronto, ON, M5S3H4, Canada}

\author[0000-0003-2377-9574]{Todd A. Thompson}
\affil{Department of Astronomy, The Ohio State University, 140 West 18th Avenue, Columbus, OH 43210, USA}
\affil{Center for Cosmology and Astroparticle Physics, The Ohio State University, 191 W. Woodruff Avenue, Columbus, OH 43210, USA}

\begin{abstract}
In previous work we identified a population of 38 cool and luminous variable stars in the Magellanic Clouds and examined 11 in detail in order to classify them as either Thorne-\.Zytkow Objects (T\.ZOs, red supergiants with a neutron star cores) or super-AGB stars (the most massive stars that will not undergo core collapse). This population includes HV\,2112, a peculiar star previously considered in other works to be either a T\.ZO or high-mass AGB star. Here we continue this investigation, using the kinematic and radio environments and local star formation history of these stars to place constraints on the age of the progenitor systems and the presence of past supernovae. These stars are not associated with regions of recent star formation, and we find no evidence of past supernovae at their locations. Finally, we also assess the presence of heavy elements and lithium in their spectra compared to red supergiants. We find strong absorption in Li and s-process elements compared to RSGs in most of the sample, consistent with super-AGB nucleosynthesis, while HV\,2112 shows additional strong lines associated with T\.ZO nucleosynthesis. Coupled with our previous mass estimates, the results are consistent with the stars being massive ($\sim$4-6.5M$_{\odot}$) or super-AGB ($\sim$6.5-12M$_{\odot}$) stars in the thermally pulsing phase, providing crucial observations of the transition between low- and high-mass stellar populations. HV\,2112 is more ambiguous; it could either be a maximally massive sAGB star, or a T\.ZO if the minimum mass for stability extends down to $\lesssim$13 M$_\odot$. 
\end{abstract}

\section{Introduction}\label{sec:intro}

While the late stages of stellar evolution are well constrained for many stars, gaps in our knowledge remain, particularly concerning the after-effects of binary interaction and the transition point between low- and high-mass stellar evolution. While progress can be made by identifying rare classes of stars at unique stages of stellar evolution, in practice multiple plausible identities can often be ascribed to stars that show unusual properties.

One such example is the peculiar star HV\,2112 in the Small Magellanic Cloud \citep{Leavitt.H.1984,Payne-Gaposchkin.Cecilia.1966.VariablesinSMC}, with two proposed identities. First, \citet{Levesque.E.2014.HV2112disc} classify it as a candidate Thorne-\.Zytkow Object (T\.ZO), which is a Red Supergiant (RSG) with a neutron star (NS) core \citep{Thorne.K.1975.TZOprime,Thorne.K.1977.TZOstructure}. T\.ZOs are potential outcomes of binary evolution following the creation of a high mass X-ray binary (HMXB) where, unlike binary neutron star (BNS) progenitors, the common envelope is not ejected. Confirming their existence would place important constraints on the rates of BNS merger events. Second, it could be a massive AGB (mAGB, M$\lesssim$6.5M$_{\odot}$) star \citep{Beasor.E.2018.HV2112AGB} or a super-AGB (sAGB, 6.5$\lesssim$M$\lesssim$12 M$_{\odot}$) star \citep{Tout.C.2014.HV2112SAGB}, the latter of which are the most massive stars that will not undergo iron core collapse \citep{Siess.L.2010.SuperAGBEarly,Jones.S.2013.SAGBEarlyModel,Doherty.C.2017.SAGBStarsECSNE}. sAGBs bridge the low-mass/high-mass divide, and their exact mass range has crucial consequences on the number of expected core collapse supernovae (CC-SNe) and subsequent chemical enrichment of the universe. Despite representing rare end-states for very different evolutionary pathways, differentiating between sAGBs and T\.ZOs photometrically or spectroscopically is difficult. 

In \citet[][hereafter Paper I]{O'Grady.A.2020.superAGBidentification} we tackled this by establishing that there exists a broader class of stars in the Small and Large Magellanic Clouds (SMC; LMC) with photometric and variability properties similar to HV\,2112. While originally identified as having stronger than expected Li and heavy element absorption (see below), HV\,2112 is also distinguished in being luminous, very cool, and showing very high amplitude variability with a peculiar double-peak morphology. We identified 10 additional stars\footnote{11 stars were initially identified in Paper I, but one (LMC-2) was found in that study to have physical properties consistent with a RSG. It is not included in this present paper, but the naming conventions of the HLOs remains consistent with Paper I, thus the label LMC-2 is skipped.} (collectively dubbed `HV 2112-like-objects' or HLOs) which all have have surface temperatures T~$<$~4800K, luminosities $\log$(L/L$_\odot$)~$>$~4.3, variability periods P~$>$~400 days, variability amplitudes $\Delta$V~$>$~2.5 mag, and a double peak feature in their light curves. We also identify 27 additional stars (dubbed `High Amplitude Variables' or HAVs) that lack the double-peaked light curve but may still belong to the same overall class. Summary tables for the HLOs and HAVs are available in Tables 2 and C1 in Paper I, respectively.

In Paper I we used the physical and variability properties and total population numbers to assess the nature of the HLO/HAVs. Broadly, they have infrared colors similar to oxygen-rich AGB stars, do not display signs of enhanced mass loss, are not consistent with observed RSG behavior, and have inferred lifetimes of a few $\times$10$^4$ years. Their temperatures and luminosities place them in the region of the Hertzsprung-Russel diagram expected for massive and super-AGB stars, though T\.ZO stars may also fall in this range. Critically, we combined their pulsation periods and measured radii with MESA stellar evolution \citep[version 10398,][]{Paxton.B.2011.MESAPaperI,Paxton.B.2015.MESAPaperII,Paxton.B.2018.MESAPaperIII,Paxton.B.2019.MESAPaperIV} and GYRE stellar pulsation \citep{Townsend.R.2013.GYRE} models to estimate current masses for these stars of $\sim$5-14M$_{\odot}$. Details of the models are provided in \S5.2.1 and \S5.4 in Paper I. This is \emph{smaller} than the theoretical minimum mass for stable T\.ZOs of M$_{\mathrm{min}}$ $\sim$15M$_{\odot}$ \citep{Cannon.R.1993.TZOStructure,Podsiadlowski.P.1995.TZOEvolution}, but consistent with the expected range for massive and superAGB stars, with most of the HLOs tending towards higher masses.

Thus, in Paper I we favored a sAGB star identity for these stars, including HV\,2112, in large part because of the mass diagnostic. However, this property is uncertain. Modern sAGB models include many advances made in stellar evolution theory over the past decades, while no modern T\.ZO models exist. Thus, assumptions had to be made about the internal structure of T\.ZOs in order to estimate their masses. While a shift in mass large enough to elevate all the HLOs above 15M$_{\odot}$ would require far larger uncertainties on our constrained physical properties than observed, the mass range for HV\,2112 in particular was estimated to be 7.5--14M$_{\odot}$, just under M$_{\mathrm{min}}$. Therefore a T\.ZO identity is still possible should the mass be underestimated. In addition, the theoretical M$_{\mathrm{min}}$ is strongly dependent on the treatment of the mass of the NS, mixing length theory, and convection in the models, and thus the true minimum mass for T\.ZOs may be lower than 15M$_{\odot}$, as discussed in detail in \S7.2.4 in Paper I. \citep{Cannon.R.1993.TZOStructure}. In that case, mass alone would not be sufficient to distinguish a T\.ZO from the most massive sAGB stars. 

To this end, we expand our analyses to further test the true identity of these stars. Here we analyze the local kinematics, local star formation history, radio environment, and (for a subset of stars) spectroscopy of the HLO/HAV sample. Motivation for each are explained in the following paragraphs, and each section will explain the relevant expectations for T\.ZOs and sAGBs given the available models. Collectively, these analyses will further elucidate the true nature of these stars. 

\emph{Kinematic Environment} (\S\ref{sec:gaia}): As the formation of a T\.ZO requires a supernova explosion, that explosion may impart a recoil velocity to the resulting T\.ZO. In order for the NS to merge with the secondary, the binary must remain bound, which can lead to a wide range of runaway or smaller 'walkaway' velocities \citep{vandenHeuvel2000,Eldridge.J.2011.WalkawayRef,Renzo.M.2019.RunawaysandWalkaways}. We will compare the proper motions of the HLOs and HAVs to their local environments to determine if any of the sample possess kinematic properties suggestive of a supernova origin.

\emph{Local Star Formation History} (\S\ref{sec:sfh}): We will compare the local star-formation histories (SFHs) of the HLOs/HAVs using detailed maps of the star formation history of the Magellanic Clouds \citep{Harris.J.2004.SFRinSMC,Harris.J.2009.SFRinLMC} to those of younger (RSGs/HMXBs) and older populations (AGBs). Since T\.ZO progenitors are stars with M$\geq15$M$_{\odot}$, and given the short ($\sim10^{5} - 10^{6}$ yr) lifetime of the T\.ZO phase \citep{Cannon.R.1993.TZOStructure,Biehle.G.1994.TZOObservational}, we expect T\.ZOs to be associated with areas of more recent ($\sim 1\times10^{7}$ yr) star formation. For sAGB stars, the total pre-AGB lifetime is predicted to be $\sim 2$-$6\times10^{7}$ yrs \citep{Doherty.C.2017.SAGBStarsECSNE}.

\emph{Radio Environments} (\S\ref{sec:radio}): We may expect to see a supernova remnant (SNR) at the location of a T\.ZO candidate. We will use the large number of radio surveys of the Clouds available to search the locations of the HLOs/HAVs for evidence of SNRs, or of a wind-driven bubble blown by the T\.ZO progenitor system within which a SNR could be expanding \citep{Ciotti.L.1989.SNRinCavity}. We do not expect evidence of SNRs around sAGB stars, though the most massive sAGBs could have blown a wind-bubble while on the main sequence.

\emph{Spectroscopy} (\S\ref{sec:spectra}): Both T\.ZOs and sAGB stars are expected to be enhanced in lithium \citep{Cameron.a.1955.BeTransport,Cameron.A.1971.SAGBLithium} and heavy elements \citep{Cannon.R.1993.TZOStructure,Biehle.G.1994.TZOObservational,Podsiadlowski.P.1995.TZOEvolution,Lau.H.2011.sprocessinSAGB,Karakas.A.2014.SAGBrpelements}. Indeed, HV\,2112 was first identified as a T\.ZO candidate by \citet{Levesque.E.2014.HV2112disc} who reported strong absorption in Li, Rb, and Mo compared to RSGs (although \citealt{Beasor.E.2018.HV2112AGB}, when comparing to stars of similar spectral type, found only a possible Li enrichment). 
While the nucleosynthetic process in T\.ZO and sAGB stars lead to similar abundance patterns, subtle differences in both heavy element abundance patterns as well as the relative timing of Li and heavy element enhancement can lead to discernible differences. We use high resolution spectroscopy to search for any excess of heavy elements in our population.

Finally, the combined results of these analyses are discussed in \S\ref{sec:disc}.


\section{Local Kinematics}\label{sec:gaia}

Here we analyze the astrometry of the HLOs and HAVs relative to their local environments. If these stars are T\.ZOs, we might expect to find evidence of a previous supernova in their kinematic properties \citep{Thorne.K.1975.TZOprime,Leonard.P.1994.TZOKick}. As radial velocities are not available for many of the stars, we instead focus on proper motions from \emph{Gaia} Early Data Release 3 \citep[EDR3,][]{Gaia.Collaboration.2020.EDR3} to assess 2D plane-of-sky velocities.

\subsection{Kinematic Expectations}\label{sec:kin_expect}

The peculiar velocity expected for a T\.ZO system depends on its formation channel. However, with the exception of dynamical formation in a dense stellar cluster, all channels require a relatively tight binary that both remains bound and has a secondary mass $>$15 M$_\odot$ after the primary explodes to form a NS. In such systems, the runaway velocity is primarily the recoil induced by the mass lost during the supernova. The impact of a random kick imparted to the NS is small, as this impulse is distributed over the (large) mass remaining in the system \citep{Liu.W.2015.LongPXraySourceTZO}. For T\.ZOs formed from high-mass X-ray binaries that subsequently undergo common envelope evolution \citep{Taam.R.1978.TZOXRBperiods} recoil velocities range from $\sim$10-80 km s$^{-1}$ with higher mass systems falling on the upper end of this range \citep{vandenHeuvel2000}. For T\.ZOs that instead form when the natal kick leads to an eccentric NS orbit with a pericenter distance inside the radius of the secondary, the median expected runaway velocity is $\sim$75 km~s$^{-1}$ \citep{Leonard.P.1994.TZOKick}.

With distances of 62.1$\pm$1.0 kpc \citep{Graczyk.D.2014.SMCDistance} and 50.0$\pm$1.3 kpc \citep{Pietrzynski.G.2013.LMCDistance} a velocity of 80 km s$^{-1}$ would correspond to a proper motion of 0.27 and 0.34 mas yr$^{-1}$ in the SMC and LMC, respectively. The kinematics of the Magellanic Clouds in the \emph{Gaia} data have been studied extensively \citep[e.g.,][]{Gaia.Collab.2018.GaiaDR2,Gaia.Collaboration.2020.EDR3,Lennon.D.2018.RunawayMCI,Oey.M.2018.RunawayMCII,DorigoJones.J.2020.RunawayMCIII} and local trends in proper motion due to, for example, the rotation of the galaxies are evident down to levels of $\lesssim$0.1 mas yr$^{-1}$. Thus, depending on location with the Clouds, the higher runaway velocities predicted for T\.ZOs may be detectable relative to the local background stars if oriented in the plane of the sky. On the other hand, we do not expect sAGBs to have significantly different kinematics from the other stars in their local environments. The non-detection of a significant peculiar velocity would be agnostic between these two origins.   

\subsection{Selection and Properties of Local Comparison Samples}

We obtain astrometric measurements from \emph{Gaia} EDR3. For each HLO and HAV, we select all \emph{Gaia} sources within 5 arcminutes of the star with \emph{Gaia} G $<$ 18 mag. This corresponds to a physical scale of $\sim$70 and $\sim$90 pc in the LMC and SMC, respectively, chosen to be significantly smaller than the physical scale over which trends due to the galaxy rotation are evident \citep{Gaia2018}.  

\subsubsection{Removal of Foreground Sources}

To avoid contamination from foreground sources in our comparison sample, we follow a procedure similar to that outlined in Paper I, based on the methods of \citet{Gaia2018}. First, we remove all sources with  parallax over parallax error ($\pi$/$\sigma_{\pi}$)~$>$~4. Then, we compare the kinematic properties of the remaining stars to the 2D distribution of proper motions formed by a sample of $\sim$1,000,000 highly probable LMC/SMC members (see Paper I for details). We assume our distribution of proper motions can be modeled as a two-dimensional Gaussian with some unknown mean $\overrightarrow{\mu}$ = ($\bar\mu_{\alpha}$, $\bar\mu_{\delta}$) and covariance matrix $\mathbf{C} = \big[\begin{smallmatrix}
  c_{\alpha\alpha} & c_{\alpha\delta}\\
  c_{\delta\alpha} & c_{\delta\delta}
\end{smallmatrix}\big]$. 

After accounting for the individual measurement uncertainties $\sigma_{\alpha,i}$ and $\sigma_{\delta,i}$ from each object $i$, the total likelihood across all objects becomes
\begin{equation}
 \ln L = -\frac{1}{2} \sum_i (\overrightarrow{\mu}_{i} - \overrightarrow{\mu})^T (\textrm{\textbf{C}}_{\textrm{tot},i})^{-1} (\overrightarrow{\mu}_{i} - \overrightarrow{\mu}) + \ln(\textrm{det}(2\pi \textrm{\textbf{C}}_{\textrm{tot},i}))
\end{equation}
where $\overrightarrow{\mu}_{i} = (\mu_{\alpha,i}, \mu_{\delta,i})$ is the proper motion of object $i$, $\textrm{\textbf{C}}_{\textrm{tot},i} = \mathbf{C} + \big[\begin{smallmatrix}
  \sigma_{\alpha,i}^{2} & 0\\
  0 & \sigma_{\delta,i}^{2}
\end{smallmatrix}\big] $ is the effective ``total'' covariance for object $i$, $T$ is the transpose operator, and $\textrm{det}$ is the determinant.

Since larger measurement uncertainties lead to larger covariances, this approach naturally weights the contribution of individual objects based on their overall measurement uncertainties. We use \texttt{scipy.optimize.minimize} to minimize the total negative log-likelihood in order to determine the optimal 5-parameter solution. We will denote this estimated optimal mean and covariance as $\overrightarrow{\mu}_{*}$ and $\mathbf{C}_{*}$, respectively, to avoid confusion with the true unknown mean and covariance $\overrightarrow{\mu}$ and $\mathbf{C}$.

For each star $j$ in each region, we then calculate a chi-square statistic as $\chi^{2}_{\textrm{j}} =$ $(\overrightarrow{\mu}_{j}-\overrightarrow{\mu}_{\textrm{med}} )^{T}\mathrm{C}_{*}^{-1}(\overrightarrow{\mu}_{j}-\overrightarrow{\mu}_{\textrm{med}})$, where $\overrightarrow{\mu}_{j}$ is the proper motion of star $j$, $\overrightarrow{\mu}_{\textrm{med}}$ is the median proper motion of the comparison sample, and $C_{*}$ is again the optimal covariance matrix derived earlier. Sources that fall outside the region containing 99.5\% of highly probable LMC/SMC members, corresponding to $\chi^{2} > 10.6$, are removed as likely foreground stars. An average of 6.4\% of stars in each HLO or HAV region are removed.

\subsubsection{Kinematic Properties of the Local Environments}

\begin{figure*}[ht]
    \centering
    \includegraphics[width=0.95\textwidth]{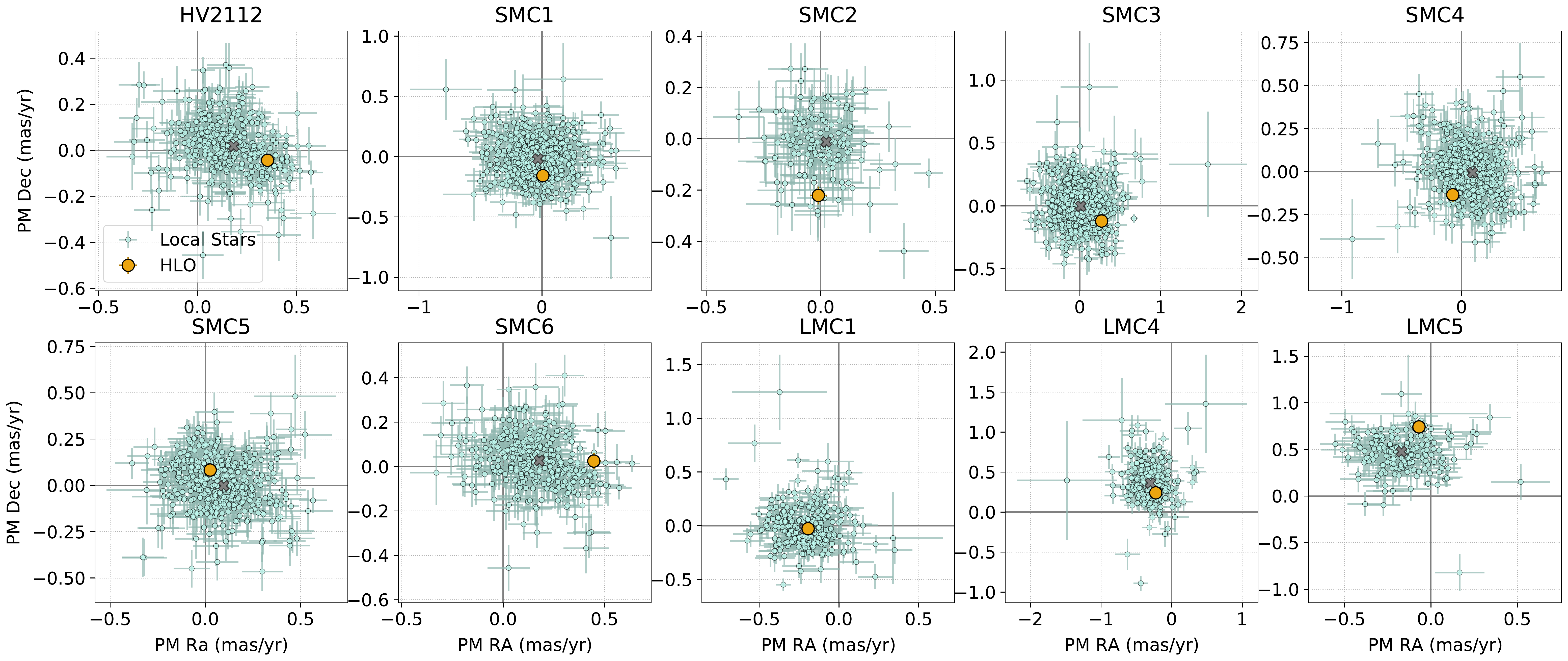}
    \caption{The proper motions of the HLOs (in gold) and all stars within 5 arc minutes that are likely Magellanic Cloud stars (in pale blue). A dark grey cross indicates the weighted mean proper motion for each group of stars. The median proper motion of the Clouds have been subtracted.}
    \label{fig:pm_sagbc}
\end{figure*}

The local comparison sample for each HLO/HAV contains approximately 500 stars. In Figure~\ref{fig:pm_sagbc}, we plot $\mu_{\alpha}$ and $\mu_{\delta}$ for the stars surrounding the 10 HLOs, after subtracting off the center-of-mass proper motions for the LMC/SMC\footnote{($\mu_{\alpha}$, $\mu_{\delta}$) = (1.89,0.31) and (0.69,$-$1.23) mas/yr for the LMC and SMC, respectively \citep{Gaia2018}.}. The weighted mean and standard deviations for the stars in each of these regions, indicated by a grey cross in Figure~\ref{fig:pm_sagbc}, are also listed in Table~\ref{tab:kinematics}. 
The fields have residual mean proper motions, relative to the center-of-mass of the Clouds, ranging from $\sim$0.0--0.3 mas yr$^{-1}$, and standard deviations of $\sim$0.1--0.15 mas yr$^{-1}$ (corresponding to velocity dispersions of $\sim$25-45 km s$^{-1}$ in the LMC and SMC).

\subsection{Comparison of HLO/HAV Kinematics to their Local Environments}

The EDR3 proper motions for the 10 HLOs are given in Table~\ref{tab:kinematics} and plotted in Figure~\ref{fig:pm_sagbc}. None of the HLOs fall completely outside of the distribution of local stars, though several (e.g. SMC-2, SMC-6) lie near the edge. Here we quantitatively assess the location of the HLOs/HAVs within 2D distribution of proper motions, and place constraints on any peculiar velocity exhibited by the stars.

All of the HLOs and HAVs have a positive and significant Gaia EDR3 `excess noise' parameter ($\epsilon_{i}$), which is a measure of how discrepant the astrometric solution of a source is from the expected model. It is expected to be close to zero for good fits. A significant excess noise could be due to instrumental errors, or intrinsic effects that affect the location of the photocenter such as binarity \citep{Lindegren.L.2012.GaiaSolution,Lindegren.L.2018.GaiaAstrometricSolution}. It has also been suggested that single-star variability may lead to increased excess noise \citep{Gandhi.P.2020.GaiaExcessNoiseVar}. We additionally assess the impact of this through the 1D distribution of total proper motions in Appendix \ref{sec:totalpm}.

\subsubsection{Location within the 2D Distribution of Local Proper Motions}

We calculate the $\chi^{2}$ statistic of each HLO and HAV compared to the 2D distribution of the proper motions of nearby stars. We follow the same procedure described above for the removal of foreground sources, but this time optimizing the covariance matrix using the $\sim$500 stars in each region, excluding the HLO/HAV itself. The resulting $\chi^{2}$ statistic encodes what fraction of the local stars have kinematics that lie interior to that of the HLO/HAV. In Table~\ref{tab:kinematics} we report the $\chi^{2}$ values and cumulative distribution function (CDF) percentage of each HLO or HAV.

None of the HLOs or HAVs fall outside the kinematic region defined by 99\% of the stars in their local environments, consistent with inspecting Figure~\ref{fig:pm_sagbc}. However, 5 HLOs (50\%) and 9 HAVs (33\%) do lie in the outer 25\% of their local distributions. To assess the significance of this, we performed an additional analysis on the 1-D distributions of total proper motions of the HLOs/HAVs. Compared to 5000 other Magellanic Cloud stars with similar levels of astrometric excess noise as the HLOs/HAVs, we see similar results. Thus, we do not see evidence that the HLOs have unusual total proper motions relative to their local mean given (i) the local dispersions and (ii) their astrometric excess noise, likely induced by their strong variability. The details of this assessment are provided in Appendix \ref{sec:totalpm}.

\begin{deluxetable*}{r|DDDDrD|rDrr}
\centering
\tablecaption{Kinematic properties of the HLOs\label{tab:kinematics}.}
\tablehead{\multicolumn{1}{r|}{HLO} & \multicolumn2r{$\mu_{\alpha}$\tablenotemark{a}} &  \multicolumn2r{$\mu_{\delta}$\tablenotemark{a}} & \multicolumn2r{$\mu_{\alpha,loc}$\tablenotemark{b}} &  \multicolumn2r{$\mu_{\delta,loc}$\tablenotemark{b}} &  \colhead{$\chi^{2}$} & \multicolumn{2}{r|}{CDF\tablenotemark{c}} & \multicolumn1r{$\mu$\tablenotemark{d}} & \multicolumn2r{CDF\tablenotemark{c}} & \colhead{Velocity\tablenotemark{e}} & \colhead{$\sigma_{Local}$\tablenotemark{f}} \\
\multicolumn{1}{r|}{} & \multicolumn2r{(mas/yr)} &  \multicolumn2r{(mas/yr)} & \multicolumn2r{(mas/yr)} &  \multicolumn2r{(mas/yr)} &  \colhead{} & \multicolumn{2}{r|}{2D} & \multicolumn1r{(mas/yr)} & \multicolumn2r{1D} & \colhead{(km/s)} & \colhead{(km/s)}
}
\decimals
\startdata
HV 2112 & 0.35$\pm$0.04 & $-$0.04$\pm$0.03 & 0.18$\pm$0.15 & 0.02$\pm$0.09 & 1.65 & 56.2\% & 0.18$\pm$0.03 & 56.0\% & 53$\pm$13 & 30  \\ 
SMC-1 & 0.01$\pm$0.04 & $-$0.16$\pm$0.03 & $-$0.03$\pm$0.14 & $-$0.02$\pm$0.11 & 1.60 & 55.0\% & 0.14$\pm$0.03 & 43.2\% & 43$\pm$12 & 39  \\ 
SMC-2 & $-$0.01$\pm$0.04 & $-$0.22$\pm$0.03 & 0.03$\pm$0.10 & $-$0.01$\pm$0.09 & 8.08  & 98.2\% & 0.21$\pm$0.03 & 76.1\% & 62$\pm$15 & 27  \\ 
SMC-3 & 0.27$\pm$0.05 & $-$0.12$\pm$0.04 & 0.01$\pm$0.16 & 0.00$\pm$0.11 & 3.05  & 78.1\% & 0.28$\pm$0.05 & 78.3\% & 83$\pm$20 & 42 \\
SMC-4 & $-$0.07$\pm$0.05 & $-$0.13$\pm$0.04 & 0.09$\pm$0.14 & $-$0.01$\pm$0.10 & 3.58  & 83.3\% & 0.21$\pm$0.04 & 67.1\% & 62$\pm$17 & 38 \\
SMC-5 & 0.03$\pm$0.03 & 0.08$\pm$0.03 & 0.10$\pm$0.15 & 0.00$\pm$0.11 & 0.59  & 25.5\% & 0.11$\pm$0.03 & 28.0\% & 33$\pm$11 & 32 \\
SMC-6 & 0.45$\pm$0.04 & 0.02$\pm$0.03 & 0.18$\pm$0.15 & 0.03$\pm$0.10 & 4.21  & 87.8\% & 0.27$\pm$0.04 & 82.8\% & 79$\pm$17 & 30 \\
LMC-1 & $-$0.19$\pm$0.04 & $-$0.03$\pm$0.04 & $-$0.22$\pm$0.11 & $-$0.03$\pm$0.13 & 0.01  & 0.5\% & 0.02$\pm$0.04 & 2.0\% & 5$\pm$9 & 34 \\
LMC-4 & $-$0.22$\pm$0.04 & 0.24$\pm$0.04 & $-$0.30$\pm$0.12 & 0.36$\pm$0.14 & 0.92  & 36.8\% & 0.14$\pm$0.04 & 41.3\% & 34$\pm$10 & 41 \\
LMC-5 & $-$0.07$\pm$0.04 & 0.74$\pm$0.05 & $-$0.17$\pm$0.14 & 0.48$\pm$0.17 & 4.38  & 88.8\% & 0.28$\pm$0.05 & 75.9\% & 67$\pm$15 & 36 \\
\hline
\hline
\enddata
\tablenotetext{a}{With the proper motion of the SMC (0.685,$-$1.230) and LMC (1.890,0.314) mas/yr \citep{Gaia.Collab.2018.GaiaDR2} subtracted}
\tablenotetext{b}{The mean weighted proper motion and standard deviation of the likely Cloud members in the 5$'$ circle around the associated HLO. Cloud proper motion has been subtracted.}
\tablenotetext{c}{Cumulative density function. The percentage denotes how much of the distribution is interior to that HLO.}
\tablenotetext{d}{The proper motion of the Cloud and the mean weighted proper motion of the local environment (columns 4 and 5) have been subtracted.}
\tablenotetext{e}{Speed relative to the average speed of stars in the local environment. Error includes proper motion measurement error, distance measurement error, and error due to Cloud depth.}
\tablenotetext{f}{Standard deviation of the proper motion of stars in the 5 arcmin bubble around each HLO.}
\end{deluxetable*}

\subsubsection{Limits on Peculiar Tangential Velocity}
 
Here, we calculate the total tangential velocity of the HLOs/HAVs relative to the local mean. We convert the total residual proper motions to velocities. The results are listed in Table~\ref{tab:kinematics} and shown on the right axes of Figure~\ref{fig:sagb_whisker}. When calculating the errors listed in Table~\ref{tab:kinematics}, we consider both the statistical errors on measurements of proper motion in RA and Dec from \emph{Gaia} and the systematic uncertainty in distance due to the line-of-site depth of the stellar distribution of the LMC and SMC (5.0 and 10.0 kpc, respectively; \citealt{Yanchulova.P.2017.MCDepths}), though the Cloud depths did not contribute significantly to the final uncertainty. 

The HLOs exhibit tangential velocities relative to their local means of 5--83 km s$^{-1}$, with over half falling below 50 km s$^{-1}$. Results are similar for the HAVs. While a peculiar velocity $>$30 km s$^{-1}$ is typically considered the criterion for classification as a ``runaway'' star \citep{Blaauw.A.1961.RunawayStars}, given (i) the velocity dispersions of $\sim$40 km s$^{-1}$ in the region around each HLO/HAV and (ii) our finding above that the residual proper motions of the HLO/HAVs are consistent with expectations for a random sample of stars with similar astrometric excess noise, we consider these velocities \emph{limits} on the tangential component of any kick imparted to the systems during an earlier phase in their evolution. In conclusion, while we do not find evidence for large (and significant) peculiar velocities, given the range of predictions for T\.ZOs outlined in \S~\ref{sec:kin_expect}, this is not particularly constraining on the origin of the systems.


\section{Local Star Formation Histories}\label{sec:sfh}

Here, we assess the relative ages of the stellar populations surrounding the HLOs/HAVs. As T\.ZOs and sAGB stars form from different mass progenitor stars, their average local SFH should differ. We use the Harris \& Zaritsky \citep[][HZ hereafter]{Harris.J.2004.SFRinSMC,Harris.J.2009.SFRinLMC} SFH maps of the LMC/SMC. Constructed using photometry of millions of individual stars from the Magellanic Cloud Photometric Survey \citep{Zaritsky.D.2002.MCPSSMC,Zaritsky.D.2004.MCPSLMC}, the HZ maps present SFHs in grids of $12'$x$12'$ boxes, corresponding to $175$x$175$ and $200$x$200$ pc in the LMC and SMC, respectively. Star formation rates within each box are presented as a function of look-back time (ranging from 4 Myr to 10 Gyr) and divided into three metallicity bins. Previous studies have used these maps to constrain the ages of populations of stars and supernova remnants \citep{Badenes.Carles.2009.StellarHistoryLMC,Williams.Benjamin.2018.ProgenMassesCCSNe,Auchettl.K.2019.CCDistributioninSMC,Sarbadhicary.Sumit.2021.ModelsOldStelPop,Diaz-Rodriguez.Mariangelly.2021.ProgMassHistCCSNe}.

\subsection{Star Formation History Expectations}\label{sec:sfh_exp}

Expectations for SFHs in the local environments surrounding T\.ZOs and sAGB stars depend on a combination of properties intrinsic to the stellar systems (e.g. progenitor mass, lifetime, natal kick) and details of the measurement methodology (e.g. physical and temporal resolution). Although the initial mass range that leads to sAGB stars is thought to be about 2-3M$_{\odot}$ wide, due to its dependence on factors such as metallicity, convective overshooting, and rotation, the initial stellar mass range that may produce sAGB stars can range from $\sim$ 6.5--12 M$_{\odot}$ \citep{Garcia-Berro.E.1994.sAGBFormation,Poelarends.A.2008.ECSneSAGBChannel,Doherty.C.2017.SAGBStarsECSNE}. The total pre-AGB lifetime is $\log(t_{\rm{years}}) \sim 7.5-7.8$ \citep{Doherty.C.2017.SAGBStarsECSNE}. No kick is expected in the prior evolution of sAGB stars, and hence they should remain in their birth environment.

For T\.ZOs, the primary of the binary system must be massive enough to explode and form a NS, and the secondary must have a mass M$_{min}$$\geq$15 M$_\odot$ at the time of the supernova, with the caveat that M$_{min}$ may be lower, as discussed in \S1. Together these constrain the mass of the primary to be $\sim$12--30 M$_{\odot}$, with the lower masses only possible for low mass ratio binaries where the secondary accretes mass during Case B mass transfer \citep[e.g.,][]{vandenHeuvel2000}. The subsequent delay-time between the supernova explosion and T\.ZO formation is expected to be small: negligible for T\.ZOs formed via a NS kick and $\lesssim10^5$ years for T\.ZOs formed from high mass X-ray binaries that enter a common envelope phase \citep{Podsiadlowski.P.1995.TZOEvolution}. Similarly, while the T\.ZO lifetime is uncertain, current estimates range from $\sim10^4-10^6$ years \citep{Cannon.R.1993.TZOStructure,Biehle.G.1994.TZOObservational,Tout.C.2014.HV2112SAGB}. Thus, the age of a T\.ZO system should be dominated by the lifetime of the primary star, and range from $\log(t_{\rm{years}}) \sim 6.7-7.2$. 

While a velocity will be imparted to a T\.ZO system by the supernova, the short delay-time and lifetime limit the distance it can travel from its birth environment. Even with the maximal peculiar velocity of $\lesssim$80 km s$^{-1}$ found in \S~\ref{sec:gaia}, a system would travel $\lesssim$80 pc in 10$^6$ years---still within a single cell in the HZ star formation history maps. Thus, we may expect that a population of T\.ZOs would be associated with areas of more recent star formation, compared to a population of sAGBs. The key to the analysis is only in part the association of the HLOs with specific HZ age bins. It is also how the clustering of the HLOs in age compares to the clustering of other stellar populations with well understood physical properties -- these differences and similarities should hold independent of any shortcomings of the HZ star formation histories. We assess this using samples of comparison stars, below.

\subsection{Method Description}\label{sec:sfr_method}

We use the HZ SFH maps to assess the ages of the stellar populations that produce the HLOs/HAVs \emph{relative to those that produce other classes of stars} (e.g. RSGs, AGBs).  We use the approach of \citet{Kochanek.C.2022.VelaPulsarSFHMethod}, which is based in turn on those of \citet{Badenes.Carles.2009.StellarHistoryLMC,Badenes.C.2015.PNeMaps}. Here, we review key aspects of this methodology.

We first combine the HZ metallicity bins into a single star formation history $s_i(j)$ for each spatial bin indexed by $j$ and age indexed by $i$. The SFH maps for the LMC and SMC used different temporal bins, so we need to analyze them separately, though we reduce the overall number by 2 by combining adjacent temporal bins. If the ``efficiency'' with which a population of stars is produced at a given age is $\epsilon_i$, then the expected number of stars associated with spatial bin $j$ is 

\begin{equation}
  e_j = \sum_i \epsilon_i s_i(j).
\end{equation}

If the actual number of stars in a spatial bin is $n_j$, the Poisson likelihood of the distribution is

\begin{equation}
   \ln L = \sum_{stars} \ln \left( { r e_j^{n_j} \over n_j! } \right)
        - \sum_{all} r e_j,
      \label{eqn:poisson}
\end{equation}

\noindent where the first sum is over the spatial bins containing stars and the second sum is over all bins.  We can discard the factorial $n_j!$ since it is just a constant contribution to the likelihood.    

If we maximize Eqn.~\ref{eqn:poisson} for the efficiencies $\epsilon_i$, the uncertainties will include the Poisson uncertainties in the number of stars. However, what we require is the maximum likelihood solution for how to distribute a fixed number of stars over the  age bins. We do this by introducing the ``renormalization factor'' $r$ into Eqn.~\ref{eqn:poisson}. Optimizing the likelihood with respect to the renormalization factor, we find that

\begin{equation}
    r = N \left[ \sum_{all} e_j \right]^{-1},
\end{equation}

\noindent where $N = \sum n_j$ is the total number of stars.  We then renormalize $\epsilon_i \rightarrow r \epsilon_i$, which also rescales $e_j \rightarrow r e_j$  so that $\sum e_j = N$. Effectively, we have converted the Poisson likelihood into the multinomial likelihood for the distribution of the $N$ stars over the age bins.  

Operationally, we are using Bayesian statistics with the likelihoods and uncertainties determined using Markov Chain Monte Carlo (MCMC) methods, with $\log e_j$ as the variables.  For each trial, the values of the $\epsilon_i$ are  renormalized before evaluating the likelihood.  To avoid having some $\log e_j \rightarrow -\infty$ we included a weak prior, further discussed in \citet{Kochanek.C.2022.VelaPulsarSFHMethod}, of

\begin{equation}
  \lambda^{-2} \sum_i \left[ \ln 
   \left ( { \epsilon_i \Delta t_{i+1} \over \epsilon_{i+1} \Delta t_i}
    \right) \right]^2,
\end{equation}

\noindent on adjacent temporal bins with $\lambda = 6.91$ in the likelihood. This is purely to avoid numerical problems and has no significant effect on the results. The number of stars formed per unit SFR is $\propto \epsilon_i \Delta t_i$ where $\Delta t_i$ is the temporal width of the bin, so the prior adds a penalty of unity to the likelihood if adjacent bins  differ in the number of stars produced per unit SFR by a factor of $1000$. The prior is just to ensure numerical stability and has no significant impact on the results. We note that this analysis does not require that a given target population is complete, but does assume that the completeness does not vary across spatial bins. This should be true for the HLO/HAV populations.

If all objects in a target population were formed at a single time in the past, then one would nominally expect the number of objects associated with that time bin to equal the total number in the population and the number of objects associated with every other temporal bin to be zero. However, since the SFHs in each spatial bin are not fully orthogonal, some spillover to other temporal bins is expected in this analysis--\emph{especially when the population being examined has a small number of objects}. 

To assess this, if we create fake stellar samples associated with particular age bins and distribute them following the HZ star formation histories, the method always identifies the correct age bin, in the sense that the distribution peaks in the correct bin. As an empirical test, we divided the OGLE IV Magellanic Cloud fundamental mode Cepheid samples \citep{Soszynski.I.2015.OGLEIVCephI,Soszynski.I.2016.OGLEIVCephII,Soszynski.I.2017.OGLEIVCephIII} into four logarithmic periods bins and used these methods to estimate their ages.  Shorter periods correspond to lower masses and hence longer main sequence lifetimes, where we used the relationships between age, mass and period from \citet{Anderson.R.2016.CepheidPeriodAge}. As seen in Figure \ref{fig:sfh_cepheids} in Appendix \ref{sfhceph}, the four period bins largely sort themselves in the correct age order except for the longest periods, where there are also very limited numbers of Cepheids.

The covariances in the SFR in this type of analysis are very strong. Therefore the published uncertainties, which are simply the diagonal elements of the matrix and not the full covariances, are not directly useful for these calculations. This is why we focus much of the discussion on comparisons between well defined stellar populations, as these differences should be little affected by the unmodelled covariances. The above tests demonstrate that this method is capable of qualitatively distinguishing between populations over the age ranges expected for T\.ZO and sAGB stars (Section~\ref{sec:sfh_exp}).

Thus we perform a qualitative comparison of the predicted ages of the HLOs and HAVs \emph{relative to a well-chosen set of comparison samples} rather than performing a detailed quantitative assessment of the resulting age distributions directly.

\begin{figure*}[t]
    \centering
    \includegraphics[width=0.95\textwidth]{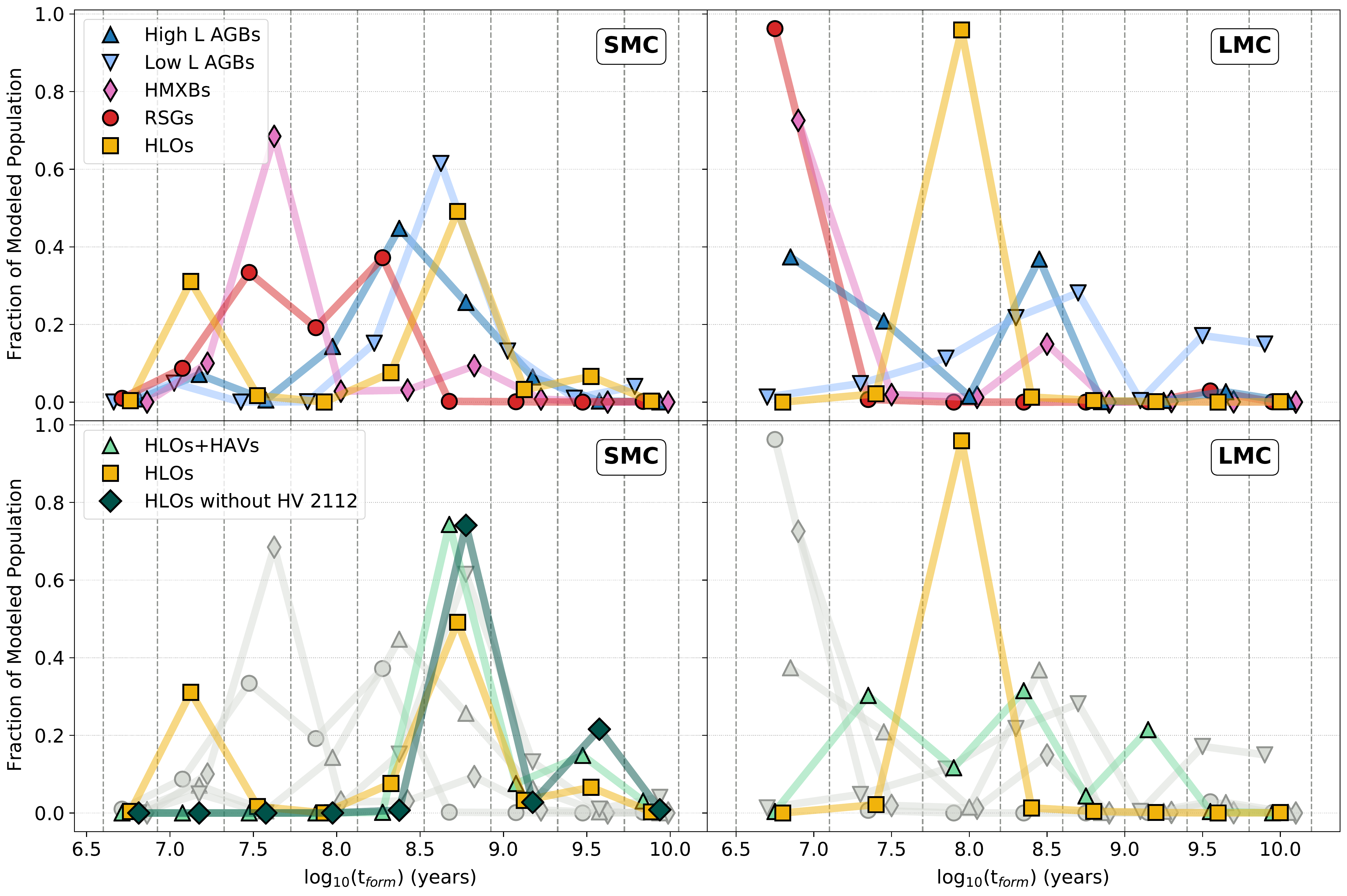}
    \caption{\textit{Top:} The fraction of each stellar population in the SMC (left) and LMC (right) located within each HZ age bin. The edges of the age bins are marked by vertical dashed lines. The HLOs are in gold, high-luminosity AGBs in dark blue, low-luminosity AGBs in light blue (see \S\ref{sfh_sample} for the distinction), RSGs in red, and HMXBs in pink. The HLOs are centered in the bins, with other populations slightly offset to aid in viewing. Uncertainties are not shown (see \S\ref{sec:sfr_method}) \textit{Bottom:} The same as above, but with the combined population of HLOs and high luminosity HAVs as light green triangles, the population of HLOs \textit{without} HV 2112 as dark green diamonds, and other populations greyed out.}
    \label{fig:sfh_hlosRSGsAGBs} 
\end{figure*}

\subsection{Comparison Sample Selection}\label{sfh_sample}

To assess the relative ages of the stellar populations in the regions surrounding the 7(3) HLOs and 3(16) high-luminosity HAVs in the SMC(LMC), we examined four comparison populations for each galaxy:

\begin{itemize}
    \item \emph{Red Supergiants} -- With estimated initial masses between $\sim$14--25 M$_\odot$ these RSGs should have similar ages to T\.ZOs. Sample Size: 28 in SMC, 104 in LMC.
    \item \emph{High Mass X-Ray Binaries} -- HMXBs with periods shorter than 100 days are predicted to be the direct ($\lesssim 10^5$ year delay-time) precursors to T\.ZOs in one of the main formation channels \citep{Taam.R.1978.TZOXRBperiods}. Sample Size: 38 in SMC, 10 in LMC.
    \item \emph{High-luminosity oxygen-rich AGB stars} -- With absolute K$_{\mathrm{s}}$ mag $<$ $-$8.95, this is expected to probe masses $\gtrsim$5 M$_\odot$, indicative of massive or super AGB stars. Sample Size: 46 in SMC, 169 in LMC.
    \item \emph{Low-luminosity oxygen-rich AGB stars} -- With expected ages older than either T\.ZO or sAGB stars, this sample provides a long-baseline control. Sample Size: 344 in SMC, 2258 in LMC.
\end{itemize}

The RSGs are from the catalogs of \citet{Massey.P.2003.RSGinMC}, \citet{Yang.M.2011.RSGsinLMC,Yang.M.2012.PLRSGSMC}, and \citet{Davies.B.2018.RSGHDL}. From \citet{Levesque.E.2005.CoolRSGs} we calculate rough masses for 123 of these 132 RSGs using the relationship $\log(M/M_\odot) = 0.5 - 0.1M_{\rm{bol}}$, which is based on fitting the Geneva evolutionary models to the mid-point of the RSG branch. Using $M_{bol} = -2.5\times\mathrm{logL_{\odot}}+4.75$ we adopt luminosities from  \citet{Neugent.K.2020.RSGBinaryLMC,Massey.P.2021.RSGWRMany}. Based on this approximate relationship, the RSGs in our comparison sample range from $\sim$14-25 M$_\odot$, with a majority between 15-20M$_{\odot}$. This is similar to the mass range expected for T\.ZOs, and indicates that our sample is not dominated by either very high or low mass RSGs.

The HMXBs come from the catalogs of \citet{Haberl.F.2016.HMXBinSMC} and \citet{Antoniou.V.2016.HMXBinLMC} for the SMC and LMC, respectively. While only $\sim$30\% of HMXBs in the Clouds have measured orbital periods, the sample with known periods is not expected to be spatially biased.

Finally, the AGBs are from \citet{Boyer.M.2011.SAGE.MC.Photom,Boyer.M.2015.SAGE.MC.Spectra}. We use only AGB stars that fall in the ``oxygen-rich'' branch, because the HLOs fall on the high-luminosity end of this sequence in the infrared color-magnitude diagram (see Paper I). For our divide between our high- and low-luminosity comparison samples, we take the K$_s$ = 9.6 magnitude of LMC-3, one of the stars originally identified in Paper I, with an estimated mass of $\sim$2$-$4M$_\odot$ based on its pulsation properties as the dividing point. It should therefore roughly separate higher mass AGB stars (M$\geq 5$M$_{\odot}$; log(t)$\lesssim$8.3) from from lower mass systems (M$\leq 4$M$_{\odot}$; log(t)$\sim$9.0).

\subsection{Results: HLOs Relative to Other Populations}

Figure~\ref{fig:sfh_hlosRSGsAGBs} shows the fraction of each population that is associated with each age bin. We do not show the bin by bin uncertainties in Figure~\ref{fig:sfh_hlosRSGsAGBs} for the reasons outlined in \S\ref{sec:sfr_method}. We do give in Table \ref{sfh:tab} (Appendix \ref{sfhtable}) the log likelihoods for the HLOs or HLOs$+$HAV populations to have the same age distribution as each of the comparison populations. These likelihoods take full account of the covariances up to the limitation discussed in \S\ref{sec:sfr_method} that we lack the covariance information for the ZH SFR histories.

In the top panels of Figure~\ref{fig:sfh_hlosRSGsAGBs} we see that the RSGs (red) are associated with young age bins. This effect is particularly strong in the LMC where the RSGs are concentrated in the youngest age bin, while in the SMC there is some blurring of the population out to $\log(t_{\rm{years}}) \sim 8.0$. However, in both galaxies the RSGs are clearly associated with younger populations than the AGB stars. HMXBs (pink) are also associated with young populations, and are the population most similar to the RSGs, as expected. In both galaxies we then observe the expected progression: high-luminosity AGBs (dark blue) are associated with older populations than RSGs/HMXBs and low-luminosity AGBs (light blue) are associated with older populations than high-luminosity AGBs. We note that some contamination between the observed samples of RSGs and high-luminosity AGBs may exist, but this should not radically change the results.

The population of HLOs is shown in gold. In the LMC (right panels), the HLOs clearly peak at formation times older than RSGs/HMXBs and slightly younger than AGB populations.  The SMC the population appears to have two peaks, one younger at earlier formation times than RSG/HMXB populations, and one older, stronger peak which is more closely associated with AGB populations. This bimodal nature of the SMC HLOs is not as conclusive as the LMC population. However, upon examining the star formation histories in the regions around the stars, we found that the region containing HV 2112 has a strong peak in star formation around $\log(t_{\rm{years}}) \sim 7.0$. If we remove HV 2112 from the SMC HLO population and repeat the analysis (dark green, bottom right panel of Figure~\ref{fig:sfh_hlosRSGsAGBs}), the younger peak disappears and the rest of the sample strongly favours ages more similar to AGB stars. 

In the bottom panels of Figure~\ref{fig:sfh_hlosRSGsAGBs}, we also show the results for the combined sample of HAVs and HLOs. In the SMC, the combined population matches the distributions of the HLOs without HV 2112 closely. In the LMC, where the population increases from 3 HLOs to 19 HLOs+HAVs, the distribution widens, but broadly peaks in the same area at the HLO population and not in the youngest bin with the RSG/HMXB populations.

If these stars were T\.ZOs, we might expect them to be consistently associated with regions of the most recent star formation, similar to their progenitor HMXBs. Thus, except for HV 2112, the SFH of the HLO/HAV population is better aligned with a sAGB star origin.


\section{Radio Environment}\label{sec:radio}

Here we search for evidence of previous supernova explosions at the locations of the HLOs and HAVs by searching for extended radio structures using the wealth of available data for the Magellanic Clouds.

\subsection{Radio Environment Expectations}

The birth of a T\.ZO requires the death of a massive star in a SN explosion, leaving behind a SNR. In the Magellanic Clouds, SNRs have been found in surveys across a range of frequencies \citep{Mathewson.D.1973.SNRinLMC,Filipovic.M.2005.ATCASMCSNRs,Williams.R.1999.SNRLMCXray,vanderHeyden.K.2004.SNRSMCXray}. Observable SNRs in the MCs have a range of physical sizes up to a cutoff at 30 pc in radius \citep{Badenes.Carles.2010.SNRinCloudsSizes}. This corresponds to an angular size of 1.6 arcmin in the SMC and 2.1 arcmin in the LMC.  Assuming a typical SNR fading time of $\sim6\times10^{4}$ yr \citep{Frail.D.1994.RadioLifetimeofSNRs,Maoz.D.2010.SNeRateinClouds}, the SNR could be visible early in the T\.ZO lifetime of $\sim10^{5} - 10^{6}$ yr \citep{Cannon.R.1993.TZOStructure,Biehle.G.1994.TZOObservational}. We have no reason to think that the selection of the HLOs and HAVs in Paper I was biased against younger T\.ZOs, though the lack of a SNR would not be conclusive evidence against a T\.ZO identity. Alternatively, the massive O-type binary progenitor system of a T\.ZO may have blown a wind bubble in the ISM. While the SNR would then not be visible until it traverses the low density medium inside the bubble, the bubble itself could be visible as an expanding shell in radio observations \citep{Ciotti.L.1989.SNRinCavity}.

Should our stars truly be sAGBs, we would not expect to see any SNRs. The most massive sAGBs ($\sim$12M$_{\odot}$) may have had strong enough winds during the main sequence lifetime to blow a wind bubble, potentially visible as a HI shell \citep{Gervais.S.1999.BubbleSearchII,Cappa.C.2000.BubbleSearch,Gaensler.Bryan.2005.StellarWindBubble}.

\subsection{Data}\label{sec_radiodata}

Here we summarize the radio data we consider:

\textit{Continuum Maps:} The spatial resolution for each map is indicated in parentheses, along with the equivalent distance in parsecs. For the SMC we use ATCA continuum images at 2.37 (0\farcm67/12pc), 4.80 (0\farcm58/11pc), and 8.64 (0\farcm37/7pc) GHz \citep{Filipovic.M.2002.ATCASMCContinuum}, and ASKAP continuum images at 960 (0\farcm5/9pc) and 1320 (0\farcm27x0\farcm25/5pc) MHz \citep{Joseph.T.2019.ASKAPEmu960and1320radio}. In the LMC we use ATCA continuum images at 4.80 (0\farcm49/9pc) and 8.64 (0\farcm37/5pc) GHz \citep{Dickel.J.2005.LMCATCAContinuum}, and ASKAP continuum images at 887 (0\farcm23x0\farcm20/3pc) MHz \citep{Harvey-Smith.L.2016.ASKAPDataProductsLMC}. We note that SNRs have been discovered and analyzed in all of these maps \citep{Filipovic.M.2005.ATCASMCSNRs,Bozzetto.L.2014.LMCSNRExample,Joseph.T.2019.ASKAPEmu960and1320radio}.

\textit{HI Velocity Cubes:} HI cubes for the SMC were taken with ASKAP \citep{DeBoer.D.2009.ASKAPPrime,McClure-Griffiths.N.2018.ASKAPHISMC}, with a spectral resolution of 4 km/s and a spatial resolution of 0\farcm58x0\farcm45 ($\sim$9 pc at the distance of the SMC). The LMC HI data were taken with ATCA \citep{Kim.Sungeun.2003.ATCAHILMCCube}, with a spectral resolution of 1.7 km/s and spatial resolution of 1$'$ ($\sim$15 pc at the distance of the SMC/LMC).

\subsection{Supernova Remnants \& Other Extended Structures}

Searching catalogs of SNRs in the Magellanic Clouds \citep{Badenes.Carles.2010.SNRinCloudsSizes,Bozzetto.L.2017.LMCSNRs,Maggi.P.2016.SNRinLMCxray,Maggi.P.2019.SNRinSMCxray}, we find none at the locations of the HLOs or HAVs. 

In the various continuum maps described in \S\ref{sec_radiodata}, we see no bubbles or other structures at the locations of any of the HLOs (except HV 2112, see \S\ref{sec_HV 2112struc}) and HAVs. For all but the archetypal HLO HV 2112, we see no evidence of any expanding shells, nor any other coherent structure, within the HI velocity cube. In the highest resolution ASKAP map, the mean 3-$\sigma$ upper limit on the surface brightness at the location of the star assuming a radius of 30 pc \citep{Badenes.Carles.2010.SNRinCloudsSizes} in the SMC is $\Sigma_{1 GHz} <$ 0.5 W m$^{-2}$ Hz$^{-1}$ sr$^{-1}$, and in the LMC is $\Sigma_{1 GHz} <$ 1.1 W m$^{-2}$ Hz$^{-1}$ sr$^{-1}$. In Appendix~\ref{hav_radio} we show the surface brightness upper limit at 30 pc for all HLOs and HAVs in Table~\ref{tab:radio_ul_all}. While we cannot rule out the existence of a very faint ($\Sigma_{1 GHz} <$ 10$^{-22}$ W m$^{-2}$ Hz$^{-1}$ sr$^{-1}$) SNR at the location of any of the stars, there is no evidence of any visible SNR for the HLOs or the HAVs.

The expected number of SNR is $N_{SNR}=n_{obj}\frac{t_{SNR}}{t_{TZO}}$ where $n_{obj}$ is 11 (HLOs) or 38 (HLOs+HAVs), $t_{SNR}$ is the fade time of 6$\times$10$^{4}$ yr, and $t_{TZO}$ is 10$^{5}$-10$^{6}$ yr. The probability of having no SNRs, $e^{-N_{SNR}}$, ranges from 0.1-51\% for HLOs only and $\sim$0-10\% for HLOs+HAVs. Unless we assume that HLOs and HAVs are separate populations (unlikely, from Paper I) and take the longest possible T\.ZO lifetime, the chance of observing no SNRs if these stars are T\.ZOs is small.

\subsection{HV 2112 Structure}\label{sec_HV 2112struc}

\begin{figure*}
    \centering
    \includegraphics[width=0.95\textwidth]{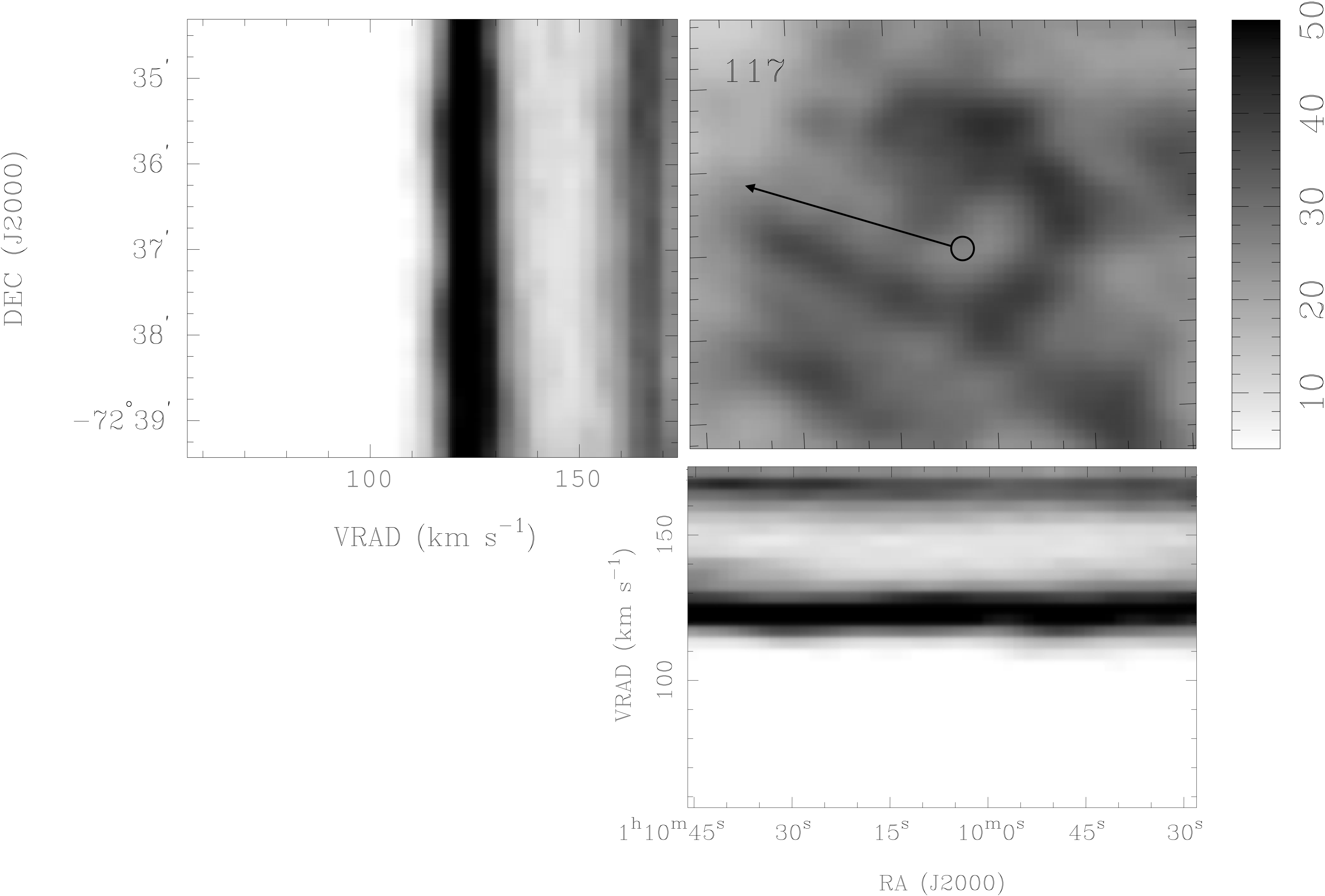}
    \caption{HI cube from ASKAP data at the location of HV\,2112. The panels show the brightness temperature in RA vs DEC at an LSR velocity of 117 km/s (top right), RA vs LSR velocity at a DEC of $-$72:36:54 (top left), and LSR velocity vs DEC at an RA of 01:10:04. The spatial resolution of the data is 0.5$'$ and the velocity resolution is 4 km/s. The position of HV\,2112 in RA vs DEC is shown with a circle, and the direction of proper motion of HV\,2112 (minus the motion of the SMC) is indicated with an arrow. At a velocity of $\sim$53km/s (Table \ref{tab:kinematics}), HV 2112 would take $\sim$1 Myr to travel the length of the arrow.}
    \label{fig:hv2112struc}
\end{figure*}

The only structure of note within the HI data at the location of any of the HLOs is a U-shaped structure surrounding HV\,2112 -- the T\.ZO candidate that defines the properties used to select the other sources -- at an LSR velocity of 116.8 km/s. The structure is shown in the top right panel of Figure~\ref{fig:hv2112struc}. It has a radius $R$ $\sim$ 27 pc (the radius of a circle that best fit the three sides of the U shape), an approximate projected area of 1890 pc$^{2}$, and a mean brightness temperature (background subtracted) of 36 $\pm$ 3 K. We do not see evidence of an expanding structure in the RA vs V and Dec vs V images,  shown in the top left and bottom right panels of Figure~\ref{fig:hv2112struc}. We therefore assume an upper limit on the expansion velocity of $v_{exp}$ $\sim$ 4 km/s, the channel width. This is less than the velocity of typical random ISM motions, so if it is not a random fluctuation, it must be a short lived structure. The HI column density through the structure is (7$\pm$2$)\times$10$^{19}$ cm$^{-2}$. If spread throughout a sphere of radius 27 pc, the average pre-existing ambient number density is n$_{o} \approx$ 0.5$\pm$0.2 cm$^{-3}$, or $\rho_{o} \approx 8\pm3 \times10^{-25}$ g cm$^{-3}$. In the interior of the structure, the mean brightness temperature is 18 $\pm$ 1 K, which gives $n_{o, int} = 0.02 \pm 0.01$ cm$^{-3}$.

We follow the same steps as \citet{Gaensler.Bryan.2005.StellarWindBubble} to investigate the possible identity of this structure.

\subsubsection{Random ISM fluctuation} The maximum expansion velocity of $v_{exp} \leq 4$ km/s is on the order of random fluctuations in the ISM ($\sim 10$km/s for an ISM temperature of $\sim 8000$K). It is therefore entirely possible that this structure is a transient, random feature in the ISM and not related to HV\,2112.

\subsubsection{Supernova Remnant}

\textit{Sedov-Taylor Adiabatic Expansion:} If this structure is a SNR in the Sedov-Taylor phase (where no heat is exchanged with the surrounding ISM), the radius and velocity are $R_{ST} = 1.54\times10^{19} (E_{51}/n_{o})^{1/5} t_{3}^{2/5} \mathrm{cm}$ and $V_{ST} = 1950 (E_{51}/n_{o})^{1/5} t_{3}^{-3/5} \mathrm{km/s}$ where $E_{51} = E/10^{51}$ ergs and $t_{3} = t/10^{3}$ yr \citep{Draine.Bruce.2011.ISMBook}. Using these equations to eliminate the age $t$ we find an explosion energy E $\sim 2.5\pm1.3 \times10^{46}$ ergs, far less than the 10$^{51}$ ergs expected for a SN. Thus this structure cannot be the Sedov-Taylor phase of a SN energy explosion.

\vspace{0.5em}

\textit{Radiative or Snowplow Phase:} If the structure is a SNR in the radiative phase (where momentum is constant), we must first find the time and radius at which the SNR left the Sedov-Taylor phase and transitioned to the radiative phase. Assuming E = $10^{51}$ ergs and using $t_{tran} = 49.3\times10^{3} E_{51}^{0.22} n_{o}^{-0.55}$ yr and $R_{tran} = 7.32\times10^{19} E_{51}^{0.29} n_{o}^{-0.42}$ cm, we find $t_{tran} = 7\pm2 \times10^{4}$ yr and $R_{tran} = 32\pm5$ pc. The observed radius of $R = 27$ pc falls within this range. In the radiative phase, $R_{Rad} = R_{tran} (t/t_{tran})^{2/7}$ and $V_{Rad} = (2/7) R_{Rad}/t$ \citep{Draine.Bruce.2011.ISMBook}. For $R_{Rad} = 27$ pc, we should see $V_{Rad} =$ 180 $\pm$ 120 km/s. We see no evidence of an expanding bubble within this velocity range in the HI data. We therefore conclude that the structure is not a SNR.

\subsubsection{Current Wind Bubble}

In order to determine if the structure could be a current wind bubble, we calculate the wind power and associated mass loss rate required to blow a bubble with $R = 27$ pc for a wind with luminosity $L_{w}$ expanding into a medium of density $\rho_{o}$. We use $R=0.76(L_{w}/\rho_{o})^{1/5}t^{3/5}$ and $v_{exp} = R/t$ \citep{Weaver.R.1977.ISMBubbles}. With a maximum $v_{exp}$ of 4 km/s, we find a minimum age of t = 4.0 $\pm$ 0.2 Myr and maximum central stellar wind power of L$_{w}$ = (7$\pm4)\times$10$^{33}$ ergs/s. The wind power is related to the mass loss rate $\dot{M}$ and wind speed by $L_{w} = \frac{1}{2}\dot{M}v_{w}^{2}$ \citep{Weaver.R.1977.ISMBubbles}. In Paper I we found that HV\,2112 had to have a $\dot{M}$ $<$ 3.8$^{+0.9}_{-0.5}$ $\times$10$^{-7}$ M$_{\odot}$/yr. For this $\dot{M}$ the wind velocity would need to be v$_{w}$ = 240$^{+70}_{-60}$ km/s. This is significantly higher than observed for AGB stars \citep[5-15 km/s,][]{Hofner.S.2018.MassLossinAGBReview}. RSGs typically have slightly higher wind velocities \citep[$\sim$ 30 km/s,][]{vanLoon.J.2005.MLRformulaRSGAGB}, and T\.ZOs are assumed to have wind speeds similar to RSGs \citep{Cannon.R.1993.TZOStructure,Biehle.G.1994.TZOObservational}. Additionally, if the structure was related to the current wind of HV\,2112, we might expect the U shape to be a bow-shock related to the motion of HV\,2112. However, the proper motion of HV\,2112, indicated in Figure~\ref{fig:hv2112struc} with an arrow, is pointing away from the bright emission arc. Thus this structure cannot be related to the current mass loss rate of HV\,2112, whether it is a super-AGB star or T\.ZO.

\subsubsection{Old Wind Bubble} Two possibilities remain, both relating to a bubble driven by a wind from a previous stage of HV\,2112's evolution. If the bubble was driven by a previous wind, the current maximum expansion velocity of $\leq$ 4 km/s could be slower than the original expansion of the wind-driven bubble.

\textit{sAGB identity:} First, if HV\,2112 is a super-AGB star with a mass of $\sim$ 8-14 M$_{\odot}$ (estimated in Paper I), its main sequence progenitor would be an early type B star. Assuming values of $\dot{M} = 1.6\times10^{-9}$ M$_{\odot}$/yr and v$_{\infty} = 4380$ km/s for a M=12.7M$_{\odot}$ B1V star \citep{Kritcka.Jiri.2014.MLRinBVs}, and therefore a wind power of L$_{w} = 9.7\times10^{33}$ erg/s, a wind-driven bubble with the ambient density and current radius would be 3.6$\pm$0.6 Myr old and have an expansion velocity of 4.5$\pm$0.7 km/s. This expansion velocity could be consistent with this structure, though we note again that we cannot observationally constrain whether the bubble is truly expanding. For a sAGB star in the mass range of 6-12M$_{\odot}$, the time since the main sequence phase (consisting mostly of the core He burning phase) would be $\sim$2-4Myr \citep{Doherty.C.2017.SAGBStarsECSNE}. Thus this structure may be a wind-blown bubble from the main sequence progenitor of a super-AGB star.

\vspace{0.5em}

\textit{T\.ZO identity:} The picture is more complex if HV\,2112 is a T\.ZO. One of the progenitor components of the T\.ZO system must be a star with M $>$ 15 M$_{\odot}$, and the other must be $\sim$12-30 M$_{\odot}$  (see \S\ref{sec:sfh_exp}). Both stars would have strong, line-driven winds on the main sequence, and could form a bubble. The later supernova would therefore not be observable as an SNR as it travels through the low density (n$_{o, int} \approx$ 0.02 cm$^{-3}$) interior of the bubble \citep{Ciotti.L.1989.SNRinCavity}. Taking two O9.5V stars with M = 16 M$_{\odot}$, $\dot{M} = 1.6\times10^{-8}$ M$_{\odot}$/yr and v$_{\infty} = 6700$ km/s \citep{Muijres.L.2012.OTypeMSMLR}, we simply sum their wind luminosities to obtain the total L$_{w}$ for the bubble \citep{Chu.You-Hua.1995.CanAddLW1,McClure-Griffiths.N.2001.CanAddLW2,Benaglia.P.2005.CanAddLW3,Ramirez-Ballina.Isidro.2019.CanAddLW4}. This gives lower limits on a total wind power of L$_{w} = 4.5\times10^{35}$ erg/s, a minimum age of 1.0$\pm$0.2 Myr, and a minimum current expansion velocity of 16$\pm$2 km/s. This age is within the predicted lifetime of a T\.ZO ($\sim10^{5} - 10^{6}$ yr \citealt{Cannon.R.1993.TZOStructure,Biehle.G.1994.TZOObservational}). However, this expansion velocity should be observable in the ASKAP HI data, and our maximum observed velocity is $\sim$4 km/s. This suggests that this structure is not the product of the main sequence winds of a T\.ZO progenitor system.

\subsubsection{Structure Identity Conclusion}

To summarize, the most likely identities of the structure around HV\,2112 is i) a random fluctuation of the ISM, or ii) a wind-driven bubble produced by the main sequence progenitor of a super-AGB star. The structure is not a supernova remnant, and could not have been created by a T\.ZO progenitor system.


\section{Spectroscopy}\label{sec:spectra}

Both T\.ZOs and m/sAGB stars are expected to synthesize high abundances of lithium and an array of heavy elements. In \S\ref{spec_expectations} we explain in detail the expectations for each class. We present our general procedure in \S\ref{spec_method} and data and reduction techniques in \S\ref{sec_obsred}. Abundance estimates for cool, inflated, stars are complex and model dependent. We use a methodology similar to \citet{Levesque.E.2014.HV2112disc}, explained in \S\ref{spec_pew_measure}-\ref{sec:temp}, where we assess the presence of potential element enhancements by comparing the ratios of absorption line equivalent widths to the same lines in a population of RSGs of similar temperatures and luminosities. Our results are reported in \S\ref{disc_specinterp}. In \S\ref{spec_levesque}-\ref{spec_beasor} we compare our results for HV\,2112 to that of \citet{Levesque.E.2014.HV2112disc} and \citet{Beasor.E.2018.HV2112AGB}, before finally interpreting the results for the entire population (\S\ref{spec_detailinterp}).

\subsection{Spectroscopic Expectations}\label{spec_expectations}

\subsubsection{T\.ZO Spectra Expectations}

Here we review the predictions for the primary nucleosynthetic outputs, although we must bear in mind that the T\.ZO models have not been updated in the decades since their introduction, and many theoretical predictions for the observed spectra of T\.ZOs were developed before several important advances in stellar evolution codes.

The luminosity of supergiant T\.ZOs comes predominantly from nuclear reactions in the hot atmosphere of the neutron star. To maintain a sufficient temperature for fusion, a minimum envelope mass of $\sim$14M$_{\odot}$ and therefore a minimum total mass of $\sim$15M$_{\odot}$ is required \citep{Cannon.R.1993.TZOStructure,Podsiadlowski.P.1995.TZOEvolution}. As these reactions take place at base of the convective envelope, fusion products can be mixed to the surface. Here we review the main nucleosynthetic outputs.

\emph{Lithium Production}: The hot temperatures and strong convection within T\.ZOs allows the synthesis of Li through the Cameron-Fowler mechanism \citep{Cameron.a.1955.BeTransport}. $^7$Be is first created and then rapidly mixed into an area with cooler temperatures where it captures an electron to form $^7$Li. Because T\.ZO lifetime constraints are comparable to the depletion timescale for $^3$He ($\sim$10$^5$ years), large quantities of Li should build up and be present at all times \citep{Podsiadlowski.P.1995.TZOEvolution}. 

\emph{The irp-process and Heavy Element Synthesis}:  
The temperature in the atmosphere of the NS in a T\.ZO is very high ($\sim$10$^9$ K), which, when coupled with the convective atmosphere, leads to the interrupted rapid proton (irp) process. Large abundances of elements such as Zn, Br, Rb, Y, Zr, Mo, and Ru will be dredged to the surface \citep{Cannon.R.1993.TZOStructure}. \citet{Biehle.G.1994.TZOObservational} found that Mo should be enhanced relative to Solar by a factor $>$1000 and Rb by a factor of $\sim$200.

\emph{Calcium Production}: Finally, \citet{Levesque.E.2014.HV2112disc} observed a higher Ca/Fe ratio in HV\,2112 than a control sample of RSGs. While Ca is not directly synthesized within a T\.ZO, Ca from the core of the companion star could be mixed upwards during the formation of the T\.ZO \citep{Tout.C.2014.HV2112SAGB}.

\emph{Summary:} Should any of the HLOs or HAVs be T\.ZOs, we would expect to see enhancements in both Li and heavy elements, with higher Mo abundances than Rb.

\subsubsection{Massive and Super-AGB Spectra Predictions}\label{sec:spec_exp_agb}

There are multiple stages in the m/s-AGB phase, which formally begins after core-helium exhaustion: (i) in the \emph{early AGB phase} stars undergo helium shell burning, (ii) eventually, the convective layer will penetrate the helium-rich zone and the \emph{second dredge-up} (2DU) will mix H-burning products to the surface and, finally (iii) there is a \emph{thermally-pulsing (TP) stage} with a thermally-unstable He-burning shell. The TP-AGB stage can be accompanied by the \emph{third dredge-up} (3DU) where He-burning products are mixed to the surface. In sAGB stars, carbon burning also ignites off-center \emph{prior} to the TP phase. There are multiple points in this evolution leading to enhanced surface abundances of various elements. Here we review the main nucleosynthetic processes and their mass dependence.

\emph{Hot Bottom Burning and Li Production:} In both massive and super-AGBs, a thin layer at the base of the convective envelope becomes so hot after the 2DU that H-burning reactions take place, a process known as Hot-Bottom Burning \citep[HBB,][]{Sackmann.I.1992.HBBandLiGiants,Garcia-Hernandez.D.2013.HBBandspinmAGBs}. This triggers the Cameron-Fowler mechanism \citep{Cameron.a.1955.BeTransport,Cameron.A.1971.SAGBLithium}, and large quantities of Li are carried to the surface. This Li is continually destroyed, and new production ceases when the supply of $^3$He has been depleted.
As a result, the period of Li enhancement is generally expected to be short-lived ($\sim10^4-10^5$ years; \citealt{Doherty.C.2014.SAGBIII}) with little to no Li remaining as the star nears the end of the TP-AGB. However, the duration of the Li rich phase increases and the time of peak Li enrichment occurs later for lower mass AGBs \citep{Garcia-Hernandez.D.2013.HBBandspinmAGBs,Karakas.A.2016.YieldsforAGBstars}. The minimum mass required to instigate HBB depends on a number of factors, including metallicity. \citet{Garcia-Hernandez.D.2007.GalacticOAGBs} predict HBB occurring for M$>$3M$_{\odot}$ at LMC metallicity, while \citet{Karakas.A.2018.AGBSMCYields} finds M$>$3.75M$_{\odot}$ at SMC metallicity.

\emph{The s-process and Heavy Element Synthesis:}
Once the TP-phase begins, 3DU events bring heavy elements, synthesized in the intershell region by the slow (s) neutron capture process, to the surface. In massive (M$>$4M$_{\odot}$) AGB stars, the primary source of free neutrons is the $^{22}$Ne($\alpha$,n)$^{25}$Mg He-burning reaction \citep{Abia.C.2002.sprocessinC,vanRaai.M.2012.RbZrLiAGBs,Karakas.A.2012.HeavyElementsBrightAGBs}. The high density, but short lifetime, of $^{22}$Ne  leads to an abundance pattern with higher production of Z$=$36-40 elements compared to heavier Z$>$40 elements (see \citealt{Doherty.C.2017.SAGBStarsECSNE} Figure 10), with a strong peak at Rubidium (Z$=$37; \citealt{Garcia-Hernandez.D.2006.RbRichAGBs}). Generally, the Rb abundance is expected to peak for mAGB stars and then decrease slightly for the higher mass  sAGB stars \citep{Karakas.A.2018.AGBSMCYields,Ritter.C.2018.sAGBYieldsetc}.

\emph{The i-process and Heavy Element Synthesis:}
In the \emph{most} massive super-AGB stars (top $\approx$0.3M$_{\odot}$), a process called \emph{dredge-out} can occur where a flux of protons is mixed into the He-burning zone near the end of the C-burning phase \citep{Ritossa.C.1999.dredgeout}. This subsequently leads to a high enough density of free neutrons (\citealt{Doherty.C.2017.SAGBStarsECSNE} \S2.3.2) to trigger the intermediate (i) neutron capture process. Although the exact i-process abundance pattern has yet to be calculated \citep{Doherty.C.2015.SAGBIV,Jones.S.2016.sAGBiprocess}, it is possible that the high neutron densities could lead to higher abundances of heavier elements (Z$>$40), such as Mo (Z=42), than the Ne$^{22}$ driven s-process described above. 

\emph{Simultaneous Li and s-process elements:}
Since the strong increase in Li after the 2DU is expected to be depleted over time, and s-process elements are only starting to build up in the thermally pulsing phase with the 3DUs, we might not expect sAGBs to show simultaneous enhancements in both. However, concurrent mild enhancement in Li and s-process elements has been observed in stars by \citet{Smith.V.1995.LiGiantsinClouds}.
\citet{Tout.C.2014.HV2112SAGB} explain that a period mild Li and s-process enhancement is possible for 10$^{4}$-10$^{5}$ years \citep{vanRaai.M.2012.RbZrLiAGBs,Doherty.C.2014.SAGBIII} during the first few thermal pulses before the Li is completely destroyed. This would be more common for lower mass AGB stars, as the period of Li enrichment lasts longer \citep{Karakas.A.2016.YieldsforAGBstars}. \citet{Mazzitelli.I.1999.LiinAGBsfullspec} also explain this phenomena as due to the 3DU already having occurred by the first thermal pulse.

\emph{Calcium Production:} Ca is not a typically expected product of nucleosynthesis in sAGBs \citep[e.g.][]{Tout.C.2014.HV2112SAGB}. It is possible that HBB could lead to the synthesis of a small amount of Ca \citep{Ventura.P.2012.CainsAGBs}, but this is poorly constrained.

\emph{Summary:}
In conclusion, AGB stars with masses above M$\gtrsim$4M$_\odot$ should be enhanced in both Li and s-process elements, with an abundance peak at Rb. The former peaks in the \emph{early} TP phase and the latter in the \emph{late} TP phase, but a brief period ($\lesssim$10$^5$ years) of moderate enhancement in both is possible, especially in lower mass AGBs. Elements heavier than Rb (i.e. Mo) will not be as strongly enhanced. Other than the potential for i-process enhancement in the \emph{most} massive sAGB stars, there is no specific nucleosynthetic signature for the off-center carbon burning that formally defines a sAGB star \citep{Doherty.C.2017.SAGBStarsECSNE}.  Constraints on mass are therefore essential to distinguishing between mAGBs and sAGBs. The mass range of sAGBs is predicted to be $\sim$6.5--12 M$_{\odot}$ \citep{Garcia-Berro.E.1994.sAGBFormation}, though at low metallicities this lower bound can extend to  $\sim$5M$_{\odot}$ \citep{Girardi.L.2000.MassTracks5SAGB,Doherty.C.2017.SAGBStarsECSNE}.

\subsection{Method and Comparison Sample}\label{spec_method}

In order to assess the possible enhancement of heavy elements and Li in the HLOs and HAVs we broadly follow the approach of \citet{Levesque.E.2014.HV2112disc} and \citet{Kuncher.M.2002.TZOSpecSearch}. The key elements we examine are Rb, Mo, and Li. We use the ratio of their pseudo-equivalent widths (p-EW) to a nearby line of Ca, K, Ni, or Fe. The ratios and rest wavelengths are shown in Table~\ref{tab:pew}, where `control' ratios are separated from `enhancement' ratios by a horizontal line. We then compare these ratios, as a function of temperature to account for both the temperature dependent nature of spectral features and the intense line blanketing effects from TiO absorption bands, to a control sample of stars in order to determine if the HLOs and HAVs show evidence of enhancements. For our control sample we chose a population of RSGs from the sample of \citet{Neugent.K.2012.RSG.YSG.LMC}, following \citet{Levesque.E.2014.HV2112disc}, since RSGs should not show enhancement in Li or s-/irp-process elements. These RSGs have temperatures and luminosities that span the range of possible HLO/HAV properties estimated in Paper I.

\begin{deluxetable}{cc}
\centering
\tablecaption{\label{tab:pew}}
\tablehead{\colhead{Pseudo-Equivalent Width Ratio} & \colhead{Short Name}}
\startdata
Li I 6707.91$\mathrm{\text{\AA}}$/Ca I 6572.78$\mathrm{\text{\AA}}$ & Li/Ca  \\ 
Li I 6707.91$\mathrm{\text{\AA}}$/K I 7698.97$\mathrm{\text{\AA}}$ & Li/K  \\ 
Mo I 5570.40$\mathrm{\text{\AA}}$/Fe I 5569.62$\mathrm{\text{\AA}}$ & Mo/Fe \\ 
Rb I 7800.2$\mathrm{\text{\AA}}$/Ni I 7797.58$\mathrm{\text{\AA}}$ & Rb/Ni \\
Rb I 7800.2$\mathrm{\text{\AA}}$/Fe I 7802.47$\mathrm{\text{\AA}}$ & Rb/Fe  \\\hline
Ni I 7797.58$\mathrm{\text{\AA}}$/Fe I 7802.47$\mathrm{\text{\AA}}$ & Ni/Fe  \\
K I 7698.97$\mathrm{\text{\AA}}$/Ca I 6572.78$\mathrm{\text{\AA}}$ & K/Ca  \\
Ca I 6572.78$\mathrm{\text{\AA}}$/Fe I 5569.62$\mathrm{\text{\AA}}$ & Ca/Fe \\
\hline
\hline
\enddata
\end{deluxetable}

\subsection{Observations and Data Reduction}\label{sec_obsred}

\subsubsection{Magellan MIKE Spectra}\label{spec_obs}

We obtained high-resolution spectroscopy of 27 stars (9 HLOs, 4 HAVs, and 14 RSGs) using the Magellan Inamori Kyocera Echelle \citep[MIKE][]{Bernstein.R.2003.MIKE} spectrograph on the 6.5-meter Magellan Clay telescope at Las Campanas Observatory, Chile. Observation dates and times are shown in Table~\ref{tab:spectra}, where HLOs, HAVs, and RSGs are separated by horizontal lines. Observations were made using 2x2 binning, `slow' readout mode, and the 1'' slit.  HV\,2112 was observed using the 0\farcs7 slit. The spectral lines listed in Table~\ref{tab:pew} are all contained within the red arm of MIKE, which has a wavelength range of $\sim$4900-9500$\mathrm{\text{\AA}}$, pixel scale of 0\farcs13/pixel, and a resolution of $R=22,000$ for the 1'' slit.

We observed all HLOs from Paper I except for SMC-4 and LMC-5, which were near the dimmest point of their light curves during our observations and hence too faint to observe (V$>$16 mag). 
The four HAVs observed were among the subset that had their physical properties measured in Paper I. Finally, we observed 14 RSGs for the control sample described in \S\ref{spec_method}, chosen to span a similar temperature range to the HLOs/HAVs (see \S~\ref{sec:temp}).

\subsubsection{Data Reduction}\label{spec_pipeline}

Initial data reduction was performed using the \texttt{CarPy} MIKE pipeline\footnote{https://code.obs.carnegiescience.edu/mike} \citep{Kelson.Dan.2000.CarPyI,Kelson.Dan.2003.CarPyII} which performs bias and flat field corrections, wavelength calibration, and extraction of individual echelle orders. The reduced data (including standard stars) is available online\footnote{ https://zenodo.org/record/7058608} on Zenodo \citep{Zenodo}. We then normalize the spectra around each feature of interest using a low order polynomial with the \texttt{IRAF} \citep{Tody.D.1986.IRAFI,Tody.D.1993.IRAFII} task \texttt{continuum}.

\begin{deluxetable*}{ccc|ccc}
\centering
\tablecaption{Spectroscopic observations\label{tab:spectra}.}
\tablehead{\colhead{Name} & \colhead{RA (J2000$^{\circ}$)} &  \multicolumn{1}{c|}{Dec (J2000$^{\circ}$)} &  \colhead{Observation Date/Time\tablenotemark{a}} & \colhead{Phase\tablenotemark{b}} & \multicolumn{1}{c}{RV ($\frac{\mathrm{km}}{\mathrm{s}}$)\tablenotemark{c}} }
\decimals
\startdata
\multicolumn{6}{c}{SMC}\\\hline
HV 2112 & 17.515856 & $-$72.614603 &  2018-03-13--23:46:01 & 0.10 & 120$\pm$6 \\ 
SMC-1 & 11.703220 & $-$72.763824 &  2020-01-18--00:27:18 & 0.00 & 184$\pm$4 \\ 
SMC-2 & 13.036803 & $-$71.606606 &  2019-12-01--00:19:47 & 0.97 & 125$\pm$3  \\ 
SMC-3 & 13.909812 & $-$73.194845 &  2019-01-01--02:55:29 & 0.21 & 188$\pm$5   \\
SMC-5 & 15.903689 & $-$73.560525 &  2019-12-01--00:33:04 & 0.15 & 176$\pm$7  \\
SMC-6 & 17.612562 & $-$72.596670 &  2019-01-01--03:09:02 & 0.85 & 127$\pm$3  \\\hline
HAV-1 & 10.339302 & $-$72.837669 &  2020-01-18--00:47:34 & 0.19 & 93$\pm$4  \\\hline
\multicolumn{6}{c}{LMC}\\\hline
LMC-1 & 80.824095 & $-$66.952095 &  2019-01-01--03:57:19 & 0.07 & 278$\pm$3  \\
LMC-3\tablenotemark{d} & 84.986223 & $-$69.589014 &  2020-01-18--01:52:56 & 0.63 & 286$\pm$2 \\
LMC-4 & 86.709478 & $-$67.246312 &  2019-01-01--04:31:36 & 0.88 & 304$\pm$2  \\\hline
HAV-2 & 74.731359 & $-$66.761542 &  2020-01-18--02:24:23 & 0.22 & 269$\pm$3  \\
HAV-3 & 81.092362 & $-$66.110381 &  2020-01-18--03:20:52 & 0.13 & 291$\pm$3  \\
HAV-4\tablenotemark{d} & 85.173783 & $-$66.246323 &  2020-01-18--05:35:12 & 0.09 & 278$\pm$13 \\\hline
RSG-1 & 73.840208 & $-$69.787972 &  2019-01-01--05:47:41 & N/A & 256$\pm$5  \\
RSG-2 & 73.883208 & $-$66.843861 &  2019-01-01--06:02:08 & '' & 299$\pm$6  \\
RSG-3 & 74.202083 & $-$69.665250 &  2019-01-01--05:37:25 & '' & 251$\pm$9 \\
RSG-4 & 76.174125 & $-$70.710333 &  2020-01-18--03:09:25 & '' & 246$\pm$6  \\
RSG-5 & 81.501417 & $-$71.596889 &  2020-01-18--04:40:39 & '' & 240$\pm$5  \\
RSG-6 & 81.947792 & $-$69.222361 &  2019-01-01--06-12-10 & '' & 274$\pm$5 \\
RSG-7 & 82.189500 & $-$69.967305 &  2019-01-01--06:23:19 & '' & 274$\pm$7 \\
RSG-8 & 82.418041 & $-$66.838028 &  2020-01-18--04:48:07 & '' & 302$\pm$4  \\
RSG-9 & 82.814917 & $-$69.066333 &  2020-01-18--04:55:35 & '' & 270$\pm$6  \\
RSG-10 & 83.209667 & $-$67.462583 &  2020-01-18--05:02:27 & '' & 290$\pm$4 \\
RSG-11 & 83.886833 & $-$69.071833 &  2020-01-18--05:10:42 & '' & 292$\pm$5  \\
RSG-12 & 84.942708 & $-$69.324471 &  2020-01-18--05:18:13 & '' & 246$\pm$6  \\
RSG-13 & 85.102000 & $-$69.354610 &  2020-01-18--05:26:28 & '' & 262$\pm$7  \\
RSG-14 & 85.271083 & $-$69.078417 &  2019-01-01--06:33:19 & '' & 253$\pm$6  \\
\hline
\hline
\enddata
\tablenotetext{a}{UTC Time at beginning of observation}
\tablenotetext{b}{Approximate phase of the variability curve the HLO or HAV was at when the observation was taken, with 0.0 = peak of light curve and 0.5 = trough.}
\tablenotetext{c}{Radial Velocity correction. See \S\ref{spec_rv} for details.}
\tablenotetext{d}{LMC-3 and HAV-4 are at lower luminosities than the other members of their respective classes, but are included for comparative purposes.}
\end{deluxetable*}

\subsubsection{Radial Velocity Determination}\label{spec_rv}

The heavy element lines we will analyze may be weak and are located in a forest of other metal lines in the spectra of these large, cool stars. To properly assess the presence of these elements, it is necessary to precisely correct our spectra to the rest frame. We carry out this process in three steps. 

First we performed a heliocentric correction on each spectrum using the \texttt{IRAF} task \texttt{rvcorrect}. Second, we perform an initial correction for the radial velocity (RV) of the stars within the MCs, computed by cross correlating our spectra with a model template. 

We measure this initial RV correction using the Ca II triplet (8498.02$\mathrm{\text{\AA}}$, 8542.09$\mathrm{\text{\AA}}$, 8662.14$\mathrm{\text{\AA}}$) and the \texttt{IRAF} package \texttt{fxcor}. For our template, we use a model spectrum from the PHOENIX library\footnote{https://phoenix.astro.physik.uni-goettingen.de/} \citep{Husser.T.2013.PHOENIXModels}. Given the cool temperatures and large radii of our stars, we used a 3200 K, log(g)=0.0 model, and corrected the model `vacuum' spectrum to `air' wavelengths. 

All of the HLOs and HAVs display reverse P Cygni emission in the Ca II triplet (shown in Figure~\ref{fig:gen_spec}; discussed in \S~\ref{spec_inv_pcyg}). This will naturally leave a small remaining shift when compared to templates with pure absorption, as the absorption component of a P Cygni feature is offset from the rest frame. Thus, after the initial RV correction was applied, we examined the Ca I 6572$\mathrm{AA}$ and Li I 6707$\mathrm{\text{\AA}}$ lines, which are isolated and strong in almost all of the HLO/HAVs. We found these lines to be offset from their expected rest wavelengths by $\sim$5--20 km/s. We therefore apply a second correction to remove these offsets. For our RSG control sample (which do not show P Cygni features) we found that any offsets of the Ca I and Li I lines relative to the Ca II RV were very small ($\sim$0-4 km s$^{-1}$). However, we applied the correction for consistency. In Table~\ref{tab:spectra}, we list the resulting RVs, which range from $93-188$ km s$^{-1}$ in the SMC and $240-304$ km s$^{-1}$ in the LMC. These values are consistent with being in the Magellanic Clouds \citep[e.g.][]{Neugent.K.2010.YSGinSMC,Neugent.K.2012.RSG.YSG.LMC,Davies.B.2018.RSGHDL}.

\subsection{Basic Spectral Features}\label{spec_inv_pcyg}

\begin{figure}
    \centering
    \includegraphics[width=0.45\textwidth]{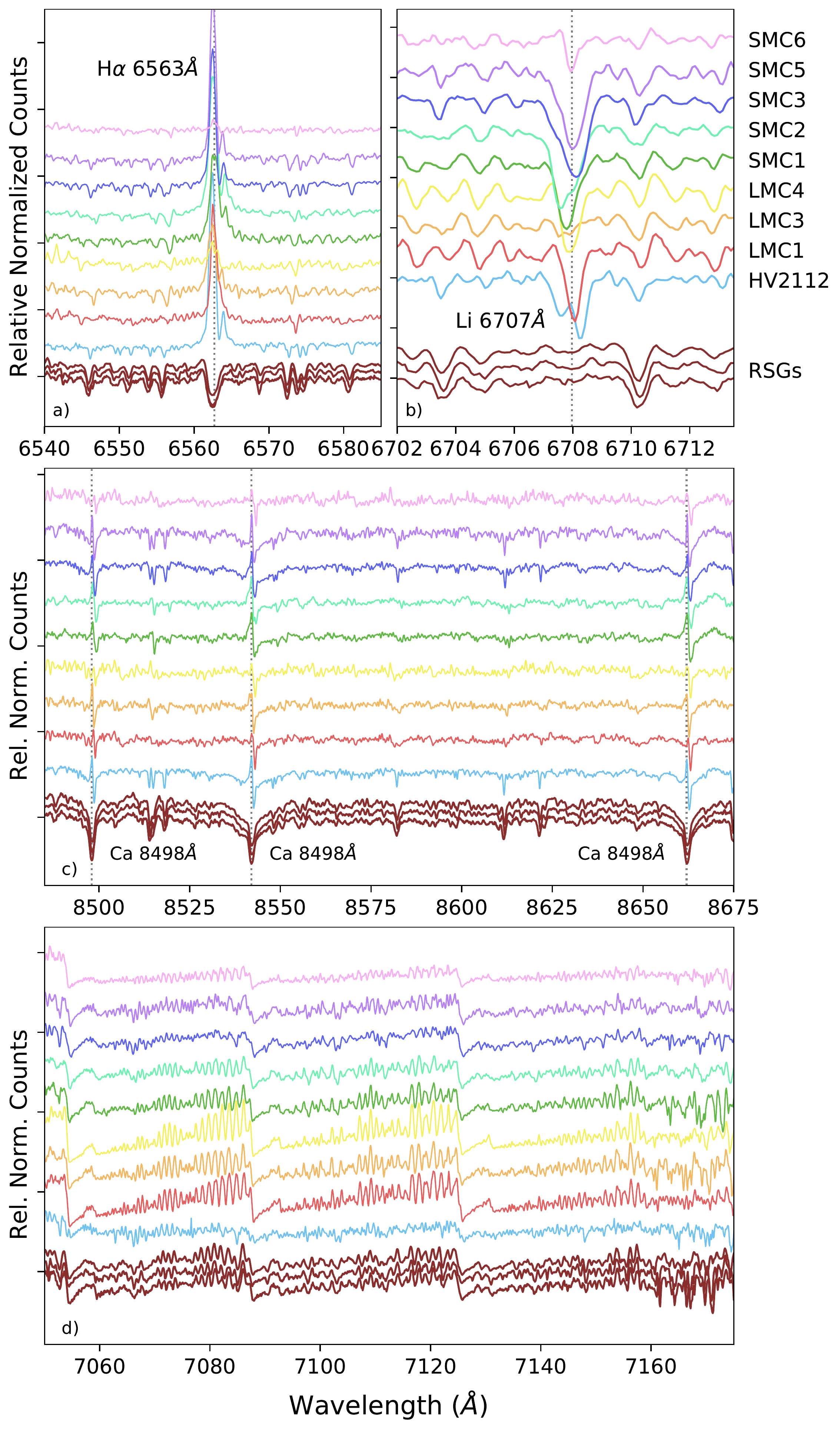}
    \caption{Spectra of all the observed HLOs and 3 RSGs (RSG-2, -8, and -13) showing various spectral features. Panel a) shows the H-$\alpha$ 6563$\text{\AA}$ feature, panel b) shows the Li 6707$\text{\AA}$ feature, panel c) shows the infrared calcium triplet, and panel d) shows an area of TiO blanketing. The HLOs are labelled on the right side of panel b).}
    \label{fig:gen_spec} 
\end{figure}

Here we present and describe the basic spectral features of the HLOs and HAVs as a population. In Figure~\ref{fig:gen_spec} we highlight several regions of interest for all the HLOs and a subset of RSGs for comparison.  Broadly, the spectra of the HLOs are consistent with expectations for cool, inflated stars. They display TiO blanketing (Fig~\ref{fig:gen_spec}; Panel D) along with a plethora of metal absorption lines. Those lines are narrow (mean Full Width at Half Maximum = 0.48$\pm$0.08$\text{\AA}$, $\approx$22km/s), indicating both a low rotation rate and low surface gravity. 

However, the spectra of the HLOs/HAVs are distinct from those of many other cool stars. They have Balmer emission with complex line profiles (Fig~\ref{fig:gen_spec}; Panel A), seen in all but HAV-1, and inverse P Cygni profiles in the Ca II triplet (Panel C), seen in all HLOs and HAVs. In contrast, the control sample of RSGs display pure absorption in both H and Ca II. Balmer emission was first noted in HV\,2112 by \citet{Levesque.E.2014.HV2112disc}.

Traditional P Cygni profiles, where the absorption is blue-shifted, are characteristic of outflows such as a stellar wind or expanding gas shell. An inverse P Cygni profile, with a red-shifted absorption component, \textit{could} be indicative of in-falling material. Indeed, inverse P Cygni profiles are common features of Young Stellar Objects \citep{Walker.M.1972.YYOri,Calvet.N.1992.BalmerTTauri,Hartmann.L.1994.TTauriRPC} and have been detected in hot LBV stars \citep{Wolf.B.1990.InversePCygLBV,Walborn.N.2017.LBVswithChangingSpectra}. Both Balmer emission and inverse P Cygni lines have also been observed in some Mira variables (cool, pulsating, giants) where they have been linked to shocks propagating through the stellar atmosphere \citep{Kudashkina.L.1994.MiraDPShocks,Barnbaum.C.1993.PCygMiraI,Richter.H.2001.PCygMiraII}

\begin{figure}
    \centering
    \includegraphics[width=0.45\textwidth]{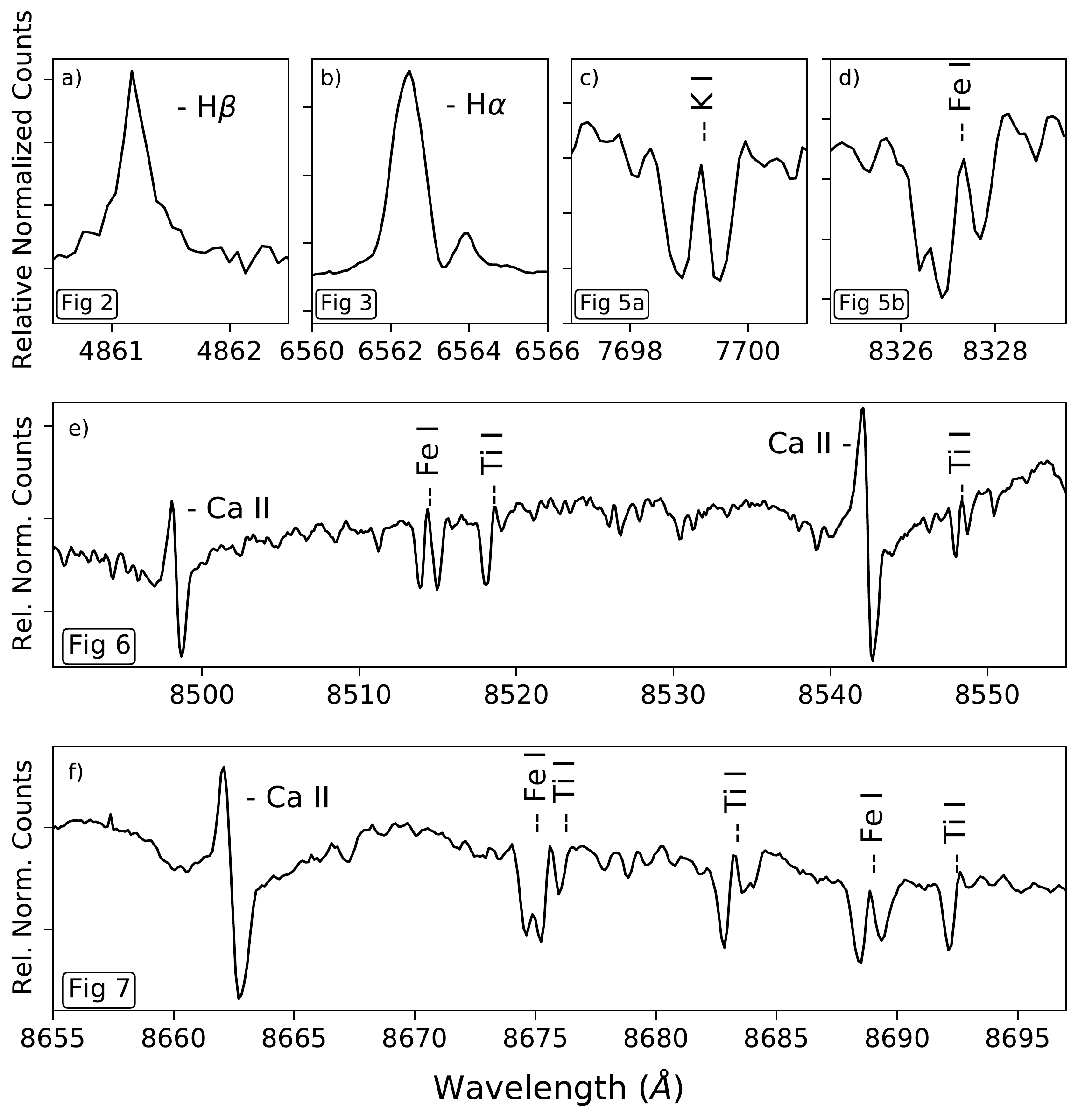}
    \caption{Spectral features of HV\,2112: The H$\beta$ line (a), the H$\alpha$ line (b), the K I doubled lines (c), the Fe I doubled lines (d), and the Ca II triplet (e and f). The annotations in the bottom left of each panel indicate the Figure number in \citet{Gillet.D.1985.SCarEmissionPRIMARY} for comparison. }
    \label{fig:scar_comparison} 
\end{figure}

In particular, we consider the Mira S Car, an AGB star with a similar effective temperature but a lower luminosity and a shorter period than HLOs/HAVs, studied in detail by \citet{Gillet.D.1985.SCarEmissionPRIMARY}. S Car displays complex Balmer features (H-$\alpha$, H-$\beta$), inverse P Cygni lines (Ca II triplet, Fe I, and Ti I), and also doubled metal lines (K I and Fe I), which are nearly identical to the features in the HLOs/HAVs (see Figures 2-6 in \citealt{Gillet.D.1985.SCarEmissionPRIMARY}). In Figure~\ref{fig:scar_comparison} we display these features for HV\,2112. \citet{Gillet.D.1985.SCarEmissionPRIMARY} argue that these features are the result of ballistic motions stemming from a passing shock. In their picture, the features are not due to in-falling material, but strong absorption lines with a central emission line. If the emission is weak, the feature appears as a double line, while if it is strong, it produces an inverse P Cygni profile. 

We therefore posit that the features described in this section are analogous to those in \citet{Gillet.D.1985.SCarEmissionPRIMARY} and are not indicative of in-falling matter, but of complex motions in the photosphere of the stars due to pulsation-driven shock waves. This is consistent with the high amplitudes of the light curves for these objects. In particular, these complex emission features are expected near photometric maximum phase, which matches the timing of our spectra, though spectra taken near the photometric minimum would be required to confirm this. The light curves of the HLOs also have bumps in the ascending phase  (Paper I Figure 4), which have also been linked to shocks \citep{Kudashkina.L.1994.MiraDPShocks}. While the HAVs do not explicitly show this double-peaked feature in their light curves (Paper I Figures 17-18), they may just be too weak or faint to see in the ASAS-SN (All-Sky Automated Survey for SuperNova) data. The lack of a double peaked light curve also does not specifically exclude the presence of shocks.

\subsection{Pseudo-Equivalent Width Measurements}\label{spec_pew_measure}

To assess the presence of heavy elements in the spectra of the HLOs/HAVs we measured the p-EWs for the features listed in Table~\ref{tab:pew}. Here we describe our procedure and details for specific lines.

We fit a Gaussian (or multiple Gaussians in the case of blended lines) to each line of interest. The local continuum level was determined by taking the mean of the data surrounding the line after sigma clipping our normalized data. Initial guesses for the location, depth, and width of the Gaussian(s) were estimated, and then allowed to vary within constraints ($\pm$0.1$\text{\AA}$ for location, $\pm$0.1 normalized flux for depth, and a maximum width of 0.25$\text{\AA}$). Line locations were taken from the Atomic Line List \citep[AtLL,][]{vanHoof.P.2018.AtomicLineList} and the NIST Atomic Spectral Database \citep[NIST ASD,][]{NIST_ASD}. 

To determine the errors on our measurements, we use a Monte Carlo approach. We generate 100 versions of each spectral segment  by allowing each point to vary within its measurement uncertainty, and repeat the continuum determination and Gaussian fitting procedure.  From this, we have 100 measurements of the equivalent width for each of the lines of interest. We report the mean of this distribution as our p-EW value, and its standard deviation as the uncertainty. These values are given in Table \ref{tab:pew_t_final} in Appendix \ref{sec:pew_t_final}. Details on specific line measurements are outlined in Appendix \ref{sec:spec_line_note}.

\subsection{Temperature Measurements}\label{sec:temp}

We want to compute the equivalent width ratios as a function of stellar temperature. 
For our control sample of RSGs, we simply use temperatures from the source catalog \citep{Neugent.K.2012.RSG.YSG.LMC}. 

For the HLOs and HAVs, we must account for the fact that their temperatures vary by $\sim$300 K throughout their pulsation periods (Paper I). We first estimate the light curve phase at the epoch when the spectra were obtained by extrapolating the period of variability and peak of the ASAS-SN light curves presented in Paper I. These extrapolated phases are listed in Table~\ref{tab:spectra}, where a phase of 0.0 is the peak of the folded light curve, which corresponds to the hottest temperature, and a phase of 0.5 is the dimmest point, which corresponds to the coolest temperature. Excluding LMC-3, all spectroscopic observations were taken within the first quarter (0.0-0.25) or last quarter (0.75-0.99) of the variability cycle. We estimate the effective temperature at the time of the spectroscopic observation by linearly interpolating between the light curve phases based on the temperatures from Paper I. For each star we adopt a generic systematic uncertainty for these temperatures by taking the average of all temperatures in either the first or last quarter of variability, corresponding to the approximate phase at the time of observation (Table \ref{tab:spectra}). These values are given in Table \ref{tab:pew_t_final} in Appendix \ref{sec:pew_t_final}.

\subsection{Element Ratio Results}\label{disc_specinterp}

\begin{figure}
    \centering
    \includegraphics[width=0.4\textwidth]{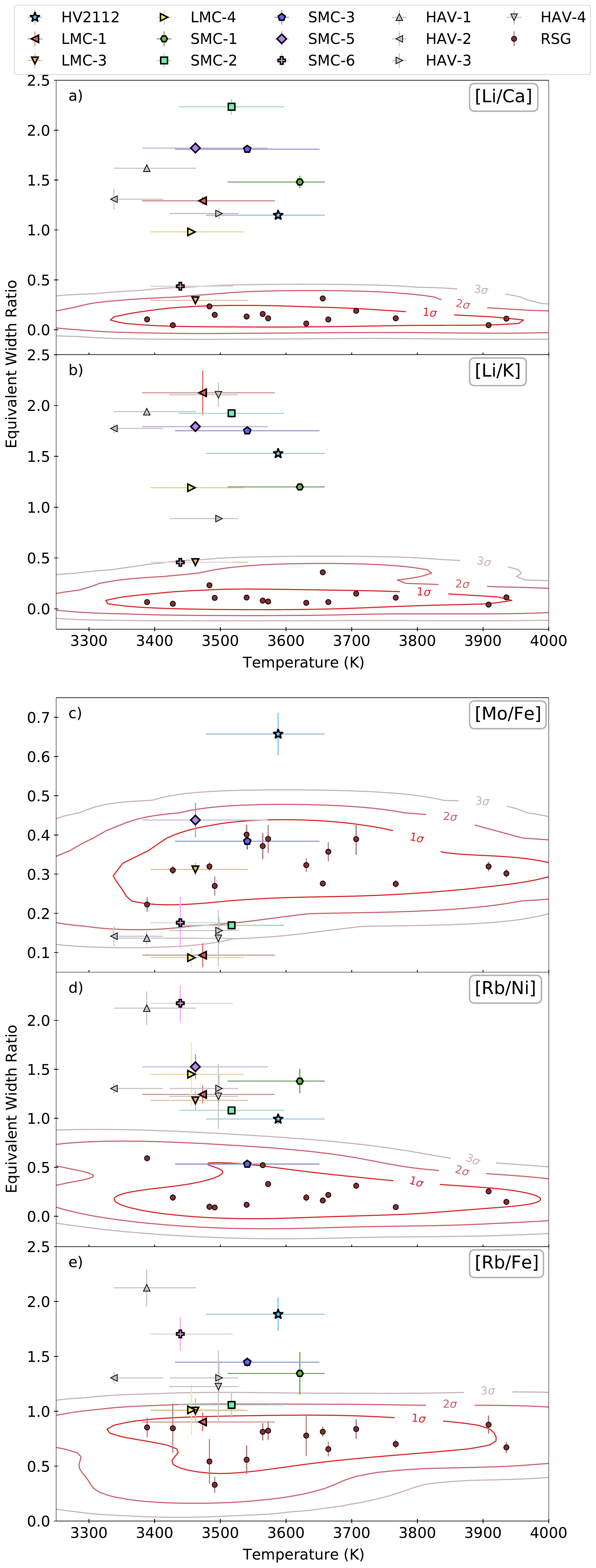}
    \caption{Pseudo-equivalent width ratios as a function of effective temperature for Li/Ca (a), Li/K (b), Mo/Fe (c), Rb/Ni (d), and Rb/Fe (e). The HLOs are shown with a variety of colors and markers as indicated in the legend. HAVs are shown in grey with different markers. RSGs are shown as dark red circles. The red lines are the 1-, 2-, and 3-$\sigma$ contours for the distribution of RSGs modeled with a kernel density estimator. Any star not appearing in a panel is explained in the text.}
    \label{fig:spec_enh_equivalent_width_ratios} 
\end{figure}

\begin{figure}
    \centering
    \includegraphics[width=0.4\textwidth]{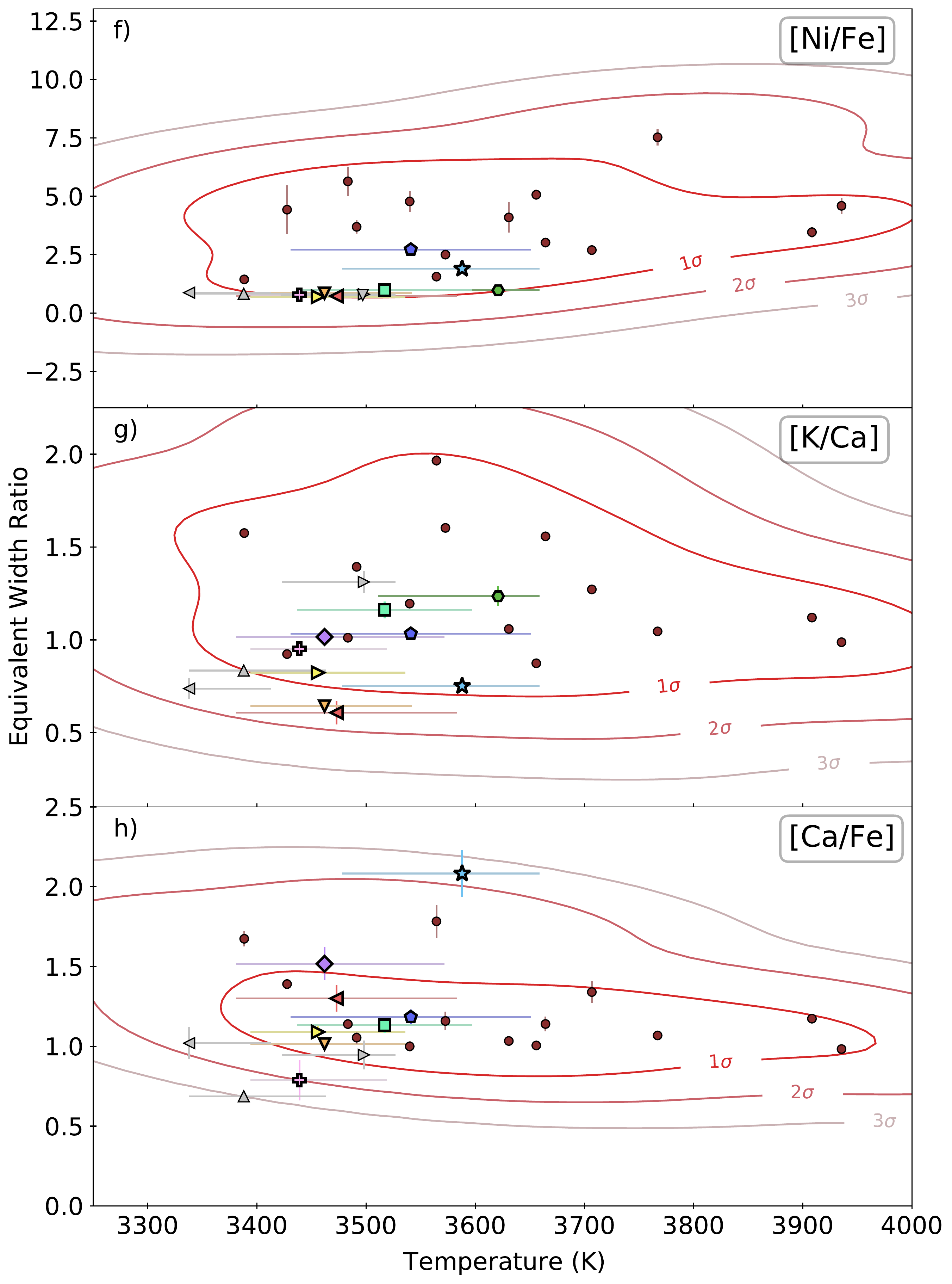}
    \caption{The same as Figure~\ref{fig:spec_enh_equivalent_width_ratios}, but for the control ratios of Ni/Fe (f), K/Ca (g), and Ca/Fe (h). }
    \label{fig:spec_con_equivalent_width_ratios} 
\end{figure}

In Figures~\ref{fig:spec_enh_equivalent_width_ratios} and~\ref{fig:spec_con_equivalent_width_ratios} we show the ratios of the p-EWs as a function of effective temperature for the HLOs, HAVs, and RSGs. Figure~\ref{fig:spec_enh_equivalent_width_ratios} shows ratios containing Li, Mo, and Rb, which should be enhanced in both T\.ZOs and sAGB stars relative to RSGs, though Mo is expected to be less enhanced than Rb for sAGBs, and Figure~\ref{fig:spec_con_equivalent_width_ratios} shows the control ratios that do not contain these elements. In all panels, contours show the \mbox{1-,} \mbox{2-,} and 3-$\sigma$ regions of the distribution of the RSG control sample. The density distribution was estimated by applying a kernel density estimator to the RSG sample. We first summarize these results for the HLOs and HAVs as a population and then compare our results for HV\,2112 to those of \citet{Levesque.E.2014.HV2112disc} and \citet{Beasor.E.2018.HV2112AGB}. 

\subsubsection{Broad Trends Among the HLO/HAV Population}\label{spec_broadtrends}

We consider a star to have a ``clearly strong line'' in absorption for a given element -- relative to the control sample -- if the equivalent width ratio falls above the 3-$\sigma$ contour of the RSG control sample. For both lithium and rubidium we consider two different line ratios (Table~\ref{tab:pew}). We refer to the potential enhancement of these elements as ``ambiguous'' if it falls above the 3-$\sigma$ threshold in one of these ratios but not the other. Evaluating these plots generally, the HLOs and HAVs fall into four broad categories, which we will relate to theoretical expectations in \S\ref{spec_detailinterp}:

\begin{itemize}
    \item Stars with clearly strong lines in Li, Rb, and Mo -- HV\,2112 only
    \item Stars with clearly strong lines in Li and Rb that lack a clearly strong Mo line -- SMC-1, SMC-5, all 4 HAVs
    \item Stars with a clearly strong Li line and ambiguous Rb line that lack a clearly strong Mo line -- SMC-2, SMC-3, LMC-1, LMC-4
    \item Stars with either a clearly strong or ambiguous Rb line that lack a clearly strong line in either Li or Mo -- SMC-6, LMC-3.
\end{itemize}

The majority of the HLOs and HAVs clearly show increases in Li/Ca and Li/K compared to the RSG control sample in Figure~\ref{fig:spec_enh_equivalent_width_ratios} (panels a and b). For ratios containing heavy elements, the results are more varied. The only star to show an excess in Mo/Fe relative to the RSG sample is HV\,2112 (panel c). Most of the HLOs and HAVs also show an increase in Rb ratios compared to RSGs (panels d and e), but some do not exceed the 3$\sigma$ level in both the Rb/Fe and Rb/Ni ratios. For the control elements show in Figure~\ref{fig:spec_con_equivalent_width_ratios}, all of the HLOs and HAVs show similar p-EW ratios to the RSGs. The only exception is that HV\,2112 shows a slight (between 2- and 3-$\sigma$) Ca/Fe excess. These results are summarized in Table~\ref{tab:spec_pew}.

The control sample RSGs are all located in the LMC, while the HLOs/HAVs are located throughout both Clouds. To investigate any bias this could introduce, we visually compare the location of HV\,2112 relative to the control sample of RSGs both in our Figures~\ref{fig:spec_enh_equivalent_width_ratios}-\ref{fig:spec_con_equivalent_width_ratios} and in Figure 1 of \citet[][who compare to SMC RSGs]{Levesque.E.2014.HV2112disc}. No strong differences can be seen. We additionally do not see any systematic differences between the LMC and SMC HLO/HAV stars in our analysis.

There are three cases where a star has not been included in a panel of Figures~\ref{fig:spec_enh_equivalent_width_ratios}-\ref{fig:spec_con_equivalent_width_ratios} because difficulties in fitting the local continuum lead to very low signal-to-noise ratios for the p-EW. SMC-1 is not included in the Mo/Fe (\ref{fig:spec_enh_equivalent_width_ratios}c) and Ca/Fe (\ref{fig:spec_con_equivalent_width_ratios}h) panels due to a low signal-to-noise ratio for Fe I 5569$\mathrm{\text{\AA}}$. SMC-5 in not included in the Rb/Fe (\ref{fig:spec_enh_equivalent_width_ratios}e) and Ni/Fe (\ref{fig:spec_con_equivalent_width_ratios}f) panels due to a very low signal-to-noise ratio for Fe I 7802.47$\mathrm{\text{\AA}}$. HAV-4 is not included in the Li/Ca (\ref{fig:spec_enh_equivalent_width_ratios}a), K/Ca (\ref{fig:spec_con_equivalent_width_ratios}g), and Ca/Fe (\ref{fig:spec_con_equivalent_width_ratios}h) panels due to a low signal-to-noise ratio for Ca I 6572.78$\mathrm{\text{\AA}}$.

\subsubsection{Comparison to Levesque et al. (2014) Measurements of HV 2112}\label{spec_levesque}

\citet{Levesque.E.2014.HV2112disc} performed a similar spectroscopic analysis on HV\,2112. Some quantitative differences are apparent in our measured p-EW ratios, likely due to a combination of physical (e.g. our spectra were obtained at a slightly different light curve phase) and systematic (e.g. we use different techniques for estimating the local continuum; treatment of doublets) effects. However, our overall results are broadly consistent: we both observe significant excesses in Li/Ca and Li/K relative to the RSG control sample and more moderate (but $>$3-$\sigma$) increases in Mo/Fe and Rb/Ni. 

We highlight two areas where our results differ slightly. First, we detect a clear high ratios in \emph{both} Rb/Ni and Rb/Fe in HV\,2112, while \citet{Levesque.E.2014.HV2112disc} observe only the former---which was noted as unusual. This seems to be driven by the large Rb/Fe ratio for the RSG control sample in \citet[][$\sim$2-4 \AA]{Levesque.E.2014.HV2112disc} compared to this work ($\sim$0.25-0.75 \AA). This difference may be due to the presence of an Fe I line close to Rb I $\lambda$7800.23 \AA\, as discussed in Appendix~\ref{sec:spec_line_note}. In our analysis we fit a double Gaussian to this blended feature and attribute only the redder component to Rb. Second, while HV\,2112 shows the highest Ca/Fe ratio of any star in our sample, this translates to only a slight excess (2-3$\sigma$) compared to the RSG control sample. In contrast, \citet{Levesque.E.2014.HV2112disc} find a clear ($>$3$\sigma$) enhancement relative to their RSG sample. 

\citet{Levesque.E.2014.HV2112disc} also reported no enhancement in the Ba II 4554 \AA\ line, another product of the s-process \citep{Vanture.A.1999.TZOUaqu}. We visually inspected this line and similarly found that it was not enhanced in any of our stars relative to the RSG sample. However as mentioned in \S\ref{sec:spec_exp_agb}, elements with Z$>$40 are not expected to be as enhanced as lighter Z=36-40 elements. For the considered metallicity range, the Ba (Z=56) production due to the Ne$^{22}$ neutron source in m/sAGB stars is predicted to be very small, with values typically less than that of Mo (Z=42, see again Figure 10 of \citealt{Doherty.C.2017.SAGBStarsECSNE}).

\subsubsection{Comparison to Beasor et al. (2018) Measurements of HV 2112}\label{spec_beasor}

\begin{figure}[ht!]
    \centering
    \includegraphics[width=0.45\textwidth]{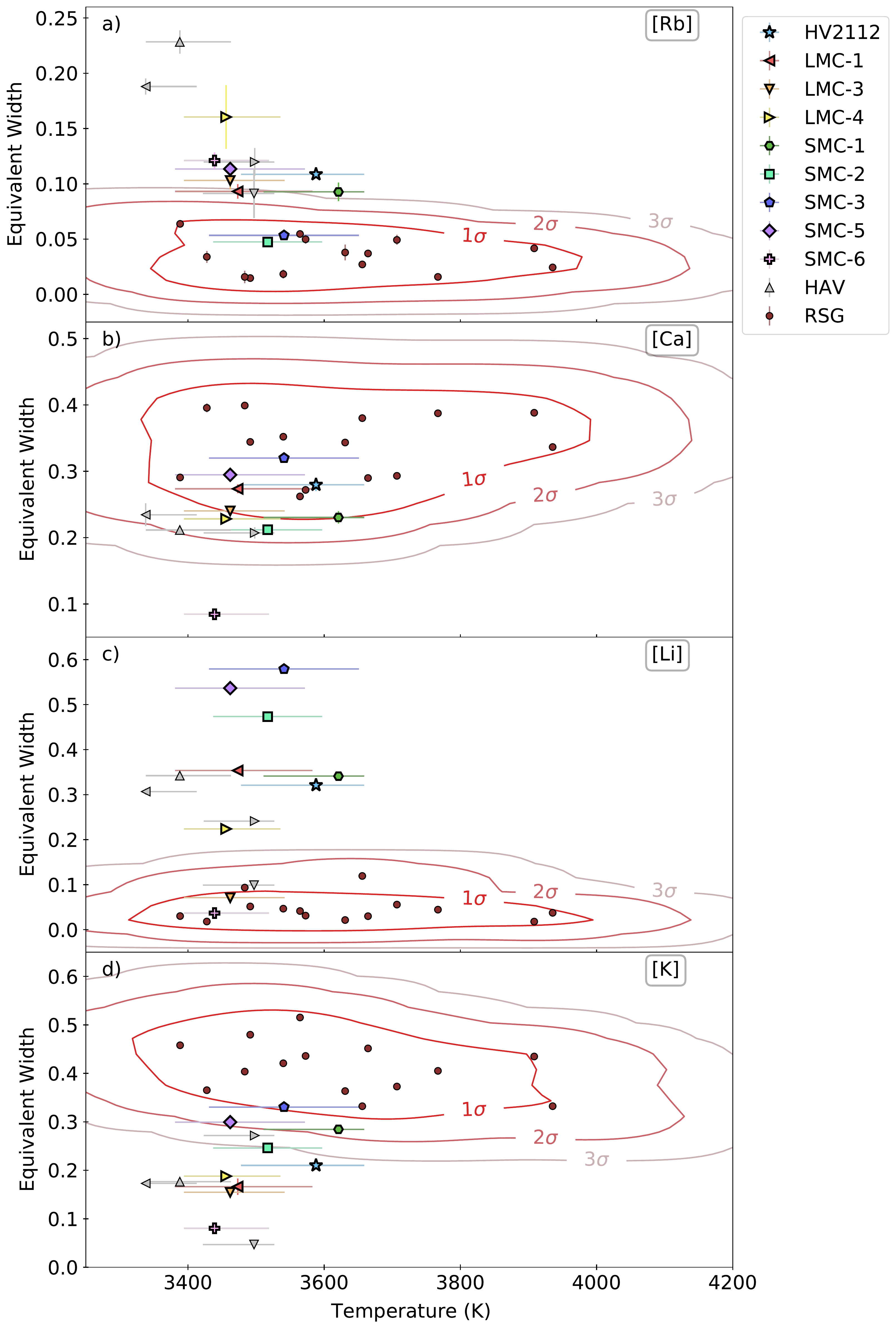}
    \caption{Plots of pseudo-equivalent width versus temperature for Rb, Ca, Li, and K. The colors, shapes, and contours are the same as Figures~\ref{fig:spec_enh_equivalent_width_ratios}-\ref{fig:spec_con_equivalent_width_ratios}. This plot can be compared with Figure 2 of \citet{Beasor.E.2018.HV2112AGB}}
    \label{fig:beasor_comp}
\end{figure}

\citet{Beasor.E.2018.HV2112AGB} also assess the spectrum of HV\,2112 and conclude that there is \emph{lack} of evidence for unusual abundances in either Rb or Ca, while a strong absorption in Li is observed and Mo is not discussed.  Examining these results, we find that our overall conclusions are actually broadly consistent. Our different ``headline'' statements can be traced primarily to our choice of comparison sample---and hence what we are assessing an ``enhancement'' relative to. 

\citet{Beasor.E.2018.HV2112AGB} select comparison stars to have as similar luminosities and spectral types to HV\,2112 as possible, as opposed to RSGs specifically. As a result, their comparison sample includes several relatively luminous AGB stars---which may themselves be enhanced in Rb and/or Li depending on their evolutionary state (see \S~\ref{sec:spec_exp_agb}). In particular, we note that of the three comparison stars that show Rb p-EWs larger than HV\,2112 in \citet{Beasor.E.2018.HV2112AGB}, two were identified has HAVs in Paper I (HV\,1719, HV\,12149) and the third (HV\,11417) would have been a candidate except that the amplitude of its V-band variability fell just below our cutoff of 2.5 mag. They are therefore members of the overall broad class of stars whose nature we are investigating. Four members of the \citet{Beasor.E.2018.HV2112AGB} comparison sample have NIR colors consistent with RSGs, based on \citet{Boyer.M.2011.SAGE.MC.Photom}. HV\,2112 does show a perceptibly higher Rb p-EW compared to those stars, although the difference is less significant than observed for Li.

As \citet{Beasor.E.2018.HV2112AGB} consider p-EW values directly, rather than ratios, we plot the Rb, Ca, Li, and K p-EWs of our entire sample (HLOs, HAVs, and RSGs) as a function of temperature in Figure~\ref{fig:beasor_comp}. Here we also see that HV\,2112 shows strong absorption in both Li and Rb lines compared to the RSG control sample, though less extreme for Rb. In both cases HV\,2112 is \emph{not} the most extreme star: other HLOs/HAVs in our sample show larger p-EWs for both elements. In our data, we find that the HV\,2112 Rb p-EW places it just above the 3$\sigma$ contours for the RSG control sample, while in \citet{Beasor.E.2018.HV2112AGB} it may be slightly less significant. 

Similar to \citet{Beasor.E.2018.HV2112AGB}, we also find that if we consider the p-EW directly, HV\,2112 does not show evidence for an excess in Ca. The 2-$\sigma$ enhancement discussed above is evident only when considered in comparison to the nearby Fe line.

\subsection{Interpretation of Element Ratio Results}\label{spec_detailinterp}

Here we interpret the results of the p-EW ratio analysis in the context of expectations for T\.ZO and sAGB stars (see \S\ref{spec_expectations}) for the four broad categories introduced in \S\ref{spec_broadtrends}. To accompany the text, Table~\ref{tab:spec_pew} lists all the p-EW ratios and whether each HLO/HAV shows a clear increase in that ratio when compared to the RSG control population. As the abundance predictions for both T\.ZO and sAGB stars in \S\ref{spec_expectations} are dependent on stellar mass, we also provide the HLO mass estimates from Paper I.

\begin{deluxetable*}{c|c||ccccc|ccc}
\centering
\tablecaption{Mass ranges \& pseudo-Equivalent Width enhancements relative to RSGs\label{tab:spec_pew}.}
\tablehead{\multicolumn{1}{c|}{Name} & \multicolumn{1}{c||}{Mass Range (M$_{\odot}$)} & \colhead{Li/Ca} & \colhead{Li/K} & \colhead{Mo/Fe} & \colhead{Rb/Ni} & \multicolumn{1}{c|}{Rb/Fe} & \colhead{Ni/Fe} & \colhead{K/Ca} & \colhead{Ca/Fe}}
\startdata
HV 2112 & 7.5--14.0 & \checkmark & \checkmark & \checkmark & \checkmark & \checkmark & X & X & $\sim$ \\ 
SMC-1 & 4.5--6.0 & \checkmark & \checkmark & N/A & \checkmark & \checkmark & X & X & N/A \\ 
SMC-2 & 4.0--10.5 & \checkmark & \checkmark & X & \checkmark & X & X & X & X \\ 
SMC-3 & 5.0-12.5 & \checkmark & \checkmark & X & X & \checkmark & X & X & X \\
SMC-5 & 2.5--6.0 & \checkmark & \checkmark & X & \checkmark & N/A & N/A & X & X \\
SMC-6 & 6.0--11.0 & $\sim$ & $\sim$ & X & \checkmark & \checkmark & X & X & X \\
LMC-1 & 5.0--11.5 & \checkmark & \checkmark & X & \checkmark & X & X & X & X \\
LMC-3 & 2.0--4.0 & X & $\sim$ & X & \checkmark & X & X & X & X \\
LMC-4 & 5.0--12.0 & \checkmark & \checkmark & X & \checkmark & X & X & X & X \\\hline
HAV-1 & N/A & \checkmark & \checkmark & X & \checkmark & \checkmark & X & X & X \\
HAV-2 & N/A & \checkmark & \checkmark & X & \checkmark & \checkmark & X & X & X \\
HAV-3 & N/A & \checkmark & \checkmark & X & \checkmark & \checkmark & X & X & X \\
HAV-4 & N/A & N/A & \checkmark & X & \checkmark & \checkmark & X & N/A & N/A \\
\hline
\hline
\enddata
\tablecomments{See Figures~\ref{fig:spec_enh_equivalent_width_ratios}-\ref{fig:spec_con_equivalent_width_ratios} for more detail. \checkmark denotes p-EW ratios higher than the 3-$\sigma$ region of the RSG control sample, $\sim$ denotes ratios between  2- and 3-$\sigma$, and X marks ratios below 2-$\sigma$. N/A denotes ratios not included due to a low signal-to-noise ratio (see text for details), and masses that were estimated in Paper I.}
\end{deluxetable*}

\emph{Strong Li, Rb, and Mo lines}: HV\,2112 shows larger p-EW ratios for all 3 elements compared to the RSGs, and it is the only star to show an enhancement in Mo/Fe. This could indicate a T\.ZO identity---increased surface abundances are expected for all of these elements in T\.ZOs. With an estimated mass of 7.5-14M$_{\odot}$, HV\,2112 also lies the closest to the theoretical minimum mass for T\.ZOs ($\sim$13-15 M$_\odot$) of any of the HLOs/HAVs.  However, Mo is expected to be enhanced by a factor of $>$5 more than Rb in T\.ZOs, which is not clear from our data. Alternatively, the mass constraints for HV\,2112 are also consistent with the most massive sAGB stars. In this case, the simultaneous Li and s-process enhancements would require HV\,2112 to be relatively early in the TP phase, before Li is significantly destroyed. Detailed abundance calculations would be required to determine if the Mo excess is consistent with an s-process abundance pattern (e.g. \citealt{Doherty.C.2017.SAGBStarsECSNE}), or if HV\,2112 is one of the \emph{most} massive sAGB stars that might undergo i-process nucleosynthesis \citep{Jones.S.2016.sAGBiprocess}. Finally, the evidence for a Ca enhancement is modest. Ca is not traditionally expected as a primary burning product in either T\.ZOs or sAGBs, although there are explanations for an enhancement for both classes of stars \citep{Tout.C.2014.HV2112SAGB,Ventura.P.2012.CainsAGBs}.

\emph{Strong Li and Rb lines}: SMC-1, SMC-5, and all 4 HAVs show clearly large Li and Rb p-EW ratios compared to RSGs, with no similar increase in Mo/Fe. This is consistent with expectations for m/sAGB stars in the TP phase. The estimated mass ranges of SMC-1 and -5 indicate that they are most likely massive rather than super AGB stars (M$_{max} \leq 6.0$M$_{\odot}$). The lack of an \emph{observable} Mo enhancement is not unexpected, as its production is predicted to be practically negligible over this mass range, and at least a factor of $\sim$3 lower than that of Rb \citep{Doherty.C.2017.SAGBStarsECSNE,Ritter.C.2018.sAGBYieldsetc,Karakas.A.2018.AGBSMCYields}. The Rb excess coupled with a lack of observed Mo is difficult to reconcile with a T\.ZO origin as Mo should be significantly enhanced in a T\.ZO.  

\emph{Strong Li and ambiguous Rb lines}: SMC-2, SMC-3, LMC-1, and LMC-4 show strong absorption in Li, but only shows a $>$3$\sigma$ enhancement in one of the two Rb ratios. If these stars are enhanced in Rb they fall into the above category, and are likely sAGBs given their masses. If they are only Li-rich, then they may be m/sAGBs in the evolutionary phase after the 2DU when Li enhancement is as its peak, but before enough 3DUs/thermal pulses have taken place to dredge up a significant quantity of s-process material. \citet{Garcia-Hernandez.D.2013.HBBandspinmAGBs} has examples of mAGBs with significant Li and little to no s-process abundance.

\emph{Strong Rb line and no Li ratio enhancement}: SMC-6 does not show a \emph{clear} excess in Li relative to the control sample. This is difficult to reconcile with a T\.ZO identity, as T\.ZOs are predicted to have strong Li enhancements throughout their entire lifetimes \citep{Podsiadlowski.P.1995.TZOEvolution}. SMC-6 is the only HLO with a minimum estimate mass \emph{above} the threshold where HBB should occur in AGB stars (M$>$3-4M$_\odot$) that does not show a significant excess in Li (although it \emph{does} display a clear excess in Rb). This elemental signature would be consistent with a more evolved TP-AGB star \citep[e.g.][]{Tabernero.H.2021.VXSagsAGB}. However, SMC-6 does \emph{not} show signatures of strong mass loss, which should also set in later in the TP phase (Paper I). It thus is possible that SMC-6 is simply at a very narrow stage in evolution where Li has been mostly destroyed, but a delayed superwind has not yet occurred \citep{Vassiliadis.E.1993.AGBEvowithMassLoss,Doherty.C.2014.SAGBMassLossCrit}. In particular, it falls very close to the 3-$\sigma$ line for the RSG control sample in both Li/Ca and Li/K in Figure~\ref{fig:spec_enh_equivalent_width_ratios}. 

\emph{Ambiguous Rb line and no Li ratio enhancement}: LMC-3 has the lowest luminosity of any of the HLOs. Its lack of Li ratio enhancement is not necessarily surprising as its estimated mass  $\sim$2-4M$_{\odot}$ falls below the range where HBB is expected to occur. However, this mass also falls below the threshold where $^{22}$Ne acts as the primary seed for the s-process, which is what leads to a heavy element abundance peaked at Rb. Thus, if LMC-3 is enriched in Rb (its line ratios were ambiguous) its mass may be underestimated and it would fall in the same narrow evolutionary phase as SMC-6.

In conclusion, to account for the possible s-process enrichment present in most of the sample, these stars must have undergone 3DU events. Most have not yet destroyed their Li enhancements. Finally, from Paper I, these stars lack strong mass loss rates ($\dot{M}\approx$10$^{-7}$M$_{\odot}$yr$^{-1}$) and have not yet reached the superwind ($\dot{M}\geq$10$^{-4}$M$_{\odot}$yr$^{-1}$) phase. Therefore, while multiple possibilities still exist for some individual stars, their spectroscopic properties paint a picture of the HLOs and HAVs being consistent with expectations for m/sAGB stars in the TP phase, but before the onset of the superwind.


\section{Implications for the Nature of the Sources}\label{sec:disc}

The goal of this work was to use 4 methods to analyze the properties of HV\,2112 and the population of similar stars identified in Paper I. Here we combine the results from all four parts of the paper: the analysis of the kinematic environment (\S\ref{sec:gaia}), the local star formation histories (\S\ref{sec:sfh}), the radio environment (\S\ref{sec:radio}), and the spectroscopic analysis of line ratios (\S\ref{sec:spectra}), in order to discuss the collective evidence for classifying them as T\.ZOs or m/sAGB stars. We discuss HV\,2112 individually in \S\ref{disc_hv2112}.

\subsection{HLOs and HAVs as a class}

Across the photometric analyses and estimation of physical properties in Paper I, the HLOs and HAVs displayed consistently similar features. This continues in the kinematic, star formation history, radio, and spectroscopic analyses. This is consistent with our hypothesis that HLOs and HAVs are one class of stars, despite the difference in light curve shape.

\subsection{Thorne-\.Zytkow Object Hypothesis}

Taken together, the results of the four analyses are difficult to reconcile with a T\.ZO origin for the HLO/HAVs as a population. On the one hand, we do not see any evidence for either a large peculiar velocity or a supernova remnant at the location of any of the HLOs/HAVs. These results do not explicitly rule out a previous SN in the evolution of the stellar system as (i) T\.ZO progenitor systems have predicted recoil velocities ranging from $\sim$10-80 km s$^{-1}$, similar to or below the general dispersion observed in the Magellanic Clouds and (ii) the radio lifetime of a typical SNR ($\sim6\times10^{4}$ yrs) \citep{Frail.D.1994.RadioLifetimeofSNRs} is shorter than the estimated T\.ZO phase lifetime of $\sim$10$^{5\mathrm{-}6}$ yrs \citep{Cannon.R.1993.TZOStructure,Biehle.G.1994.TZOObservational}, though given the size of our population, it would unlikely for us to observe zero SNRs. However, they also do not provide any specific evidence for a T\.ZO origin.

On the other hand, both the local star formation history and spectroscopic analysis pose a significant challenge for the T\.ZO hypothesis. Excepting HV 2112 (\S\ref{disc_hv2112}), as a population the HLOs/HAVs are associated with older stellar populations than $\sim$15$-$25M$_\odot$ RSGs or short-period HMXBs. While this does not rule out a T\.ZO origin for any individual object, it is difficult to reconcile with the bulk population of HLOs/HAVs being the short lived descendants of these massive stars. Similarly, while most HLOs/HAVs appear to be Li-rich, as expected for T\.ZOs, the fact that many show an observable Rb excess but all stars except HV\,2112 \emph{lack} an observable Mo excess is counter to expectations from the abundance pattern predicted for the irp-process in T\.ZOs. In particular, Mo should be enhanced by a factor of $>$1000 compared to Solar abundances and Rb by a factor of $\sim$200 \citep{Biehle.G.1994.TZOObservational}. 

We therefore conclude that the HLO/HAV population are not T\.ZOs. This is consistent with our interpretation from Paper I where we estimated their current masses to be below the $\sim$13-15M$_{\odot}$ minimum for T\.ZOs.

\subsection{Massive/super-AGB star Hypothesis}

Taken together, our results are consistent with the HLOs/HAVs being a population of massive and super-AGB stars. First, the lack of a detectable peculiar velocity or SNR is consistent with an AGB interpretation. Second, while quantitative age-dating is complicated by the low number of objects in our sample and lack of quantifiable uncertainties, the star formation history analysis demonstrates that the HLOs/HAVs are associated with stellar populations that are most similar in age to luminous AGB stars. 

Finally, as discussed in \S~\ref{spec_detailinterp} the spectroscopic features observed in the HLOs/HAVs can all be understood if they are massive (M$>$4M$_\odot$) AGB stars at various points during TP phase, prior to the activation of a superwind. We emphasize that (i) the moderate enhancement in both Li and Rb can be achieved in the early-to-mid TP phase, especially for lower mass AGB stars, (ii) the presence of a \emph{detectable} excess of Rb but not Mo is consistent with s-process nucleosynthesis in massive AGB stars, as the $^{22}$Ne($\alpha$,n)$^{25}$Mg reaction is the primary source of free neutrons and leads to an abundance pattern that peaks at Rb.

Based on the mass estimates from Paper I, the sample of HLOs and HAVs may contain both sAGB stars with M$\gtrsim$6-7M$_{\odot}$, as well as mAGB stars (e.g. SMC-1, SMC-5, and LMC-3). The HLOs and HAVs appear to be a range of evolved, intermediate mass stars, ranging from the lowest masses that achieve HBB up through the highest mass of sAGB stars in HV\,2112.

The identification of this population of m/sAGB stars has important implications to our understanding of stars near the supernovae mass boundary.  In particular, based on their number statistics, this m/sAGB sample can increase our knowledge about the duration of the pre-superwind phase and hence early TP-AGB mass-loss rates, which in turn impacts the nucleosynthesis and stellar yields.  Additionally, comparison to the observed maximum luminosities of the most massive AGB stars from this work may help constrain the very uncertain mixing length theory parameter $\alpha$ used within stellar evolutionary models during this phase.

\subsection{HV 2112}\label{disc_hv2112}

HV 2112 remains the most exceptional star in our sample. It has the highest luminosity, variability amplitude, and inferred mass of all the HLOs (Paper I). It is the only star to show an extended structure in its radio environment. HV\,2112 is the \emph{only} HLO/HAV to show strong absoprtion in Mo as well as Li and Rb. 

Many of HV\,2112's properties are consistent with \emph{both} a T\.ZO and sAGB classification. HV\,2112 shows strong lines of Li and the s-process elements compared to RSGs, expected in both types of stars. The star formation rate of its local environment peaked $\sim10^{7}$ years ago, though we cannot rule out that it was formed during an earlier, weaker burst of star formation. Finally, while we do not detect evidence of a previous supernova from the kinematic or radio environment of HV\,2112, this does not rule out a T\.ZO nature. 

There are three possible points of divergence: the estimated mass of HV\,2112, its possibly enhanced calcium abundance, and the nature of the HI structure surrounding HV\,2112. The estimated mass of HV\,2112 from paper I (7.5--14M$_{\odot}$) is slightly below the minimum stable T\.ZO mass from \citet{Cannon.R.1993.TZOStructure}. On the other hand, sAGB stars should not have a Ca enrichment, and HV\,2112 shows a small increase in Ca relative to K when compared to RSGs. Finally, while the structure surround HV\,2112 (\S\ref{sec_HV 2112struc}) is likely a random, unrelated ISM fluctuation (and thus not constraining to its origin) the possibility remains that it could be a small wind bubble blown by a $\sim$12M$_{\odot}$ progenitor of a super-AGB star, which is consistent with the estimated mass range.

Assuming the HI structure is indeed a random ISM fluctuation, then if either the current mass of HV\,2112 from Paper I is slightly underestimated \emph{or} if the minimum stable mass for T\.ZOs (which is sensitive to the treatment of convection) is actually $<$15 M$_\odot$ then HV\,2112 would remain a strong T\.ZO candidate. In particular, \citet{Biehle.G.1994.TZOObservational} speculates that the minimum mass of T\.ZOs should be $\sim$13 M$_\odot$, just inside the estimated mass range of HV\,2112. Additional modeling of T\.ZO structure and pulsation properties would enable better mass constraints and illuminate the nature of this star. Modern modeling of T\.ZOs may also revise other constraints; should updated theoretical predictions indicate longer T\.ZO lifetimes and a diversity of surface compositions, there would be no tension between our analyses and a T\.ZO identity for HV 2112.

However, if HV\,2112 either does not have an overabundance of Ca (note the discussion in \S~\ref{spec_beasor}) or if the speculation of \citet{Ventura.P.2012.CainsAGBs} that small quantities of Ca could be produced in sAGB stars is true, then all the properties of HV\,2112 are consistent with a super-AGB star near the beginning of the TP-phase. In particular the current mass estimate and observed excess in Mo would favor a sAGB at the upper end of the mass range and make it a candidate electron-capture supernova progenitor \citep{Miyaji.S.1980.OriginalecSN,Doherty.C.2017.SAGBStarsECSNE}.


\section{Summary \& Conclusions}\label{sec:concl}

\emph{Motivation}: Our goal in this paper was to clarify the nature of a population of super-AGB star candidates identified by \citet{O'Grady.A.2020.superAGBidentification}, and to see if any evidence supported a T\.ZO identity for any of the stars. We investigated their kinematics, local star formation history, and radio properties. Finally, we analyzed spectra of a subset of the sample to determine if the stars show evidence of Li and s- or irp-process enrichment.

\emph{Gaia Kinematics}: We found no evidence of hyper-velocity stars in our sample, but were unable to conclusively rule out a lower `runaway' or `walkaway' velocity for any of the stars, mostly due to uncertainties on the proper motion measurements. Compared to the mean motion and dispersion of their local environment, our stars do not display significantly different proper motions. Uncertainties on proper motion measurements are expected to improve \textit{significantly} by the end of the \textit{Gaia} mission, therefore future releases will allow tighter constraints to be placed on the motions of these stars.

\emph{Local Star Formation History}: We found that these stars are not associated with regions of recent star formation, and are more closely associated with stellar populations that will become AGB stars than those that will become RSGs.

\emph{Radio Environments}: We saw no evidence of supernova remnants at the locations of the stars. A U-shaped structure exists around HV\,2112, but it is most likely an unassociated, random fluctuation in the ISM.

\emph{Spectroscopy}: HV\,2112 displays significant Li, Rb, and Mo excesses, and a less significant Ca excess. All of the other stars in our sample have an Li excess, most have an Rb excess, and none show an excess in Mo or Ca compared to RSGs.

\emph{Population Final Conclusions}: Collectively, the analyses do not provide any evidence pointing towards a T\.ZO classification for these stars. Strong Li and Rb lines with weak Mo lines, relative to RSGs, and their estimated masses ($\sim$4-13M$_{\odot}$) strongly indicates that these stars (except for HV\,2112) are massive or super-AGB stars in the TP-AGB phase prior to the superwind onset. This would be the first population of super-AGB stars identified.

\emph{HV 2112 Final Conclusions}: Due to an estimated mass too small to be a T\.ZO, and a possible Ca enhancement which is not expected in sAGB stars, the true nature of HV\,2112 remains ambiguous. If HV\,2112 is not a T\.ZO, given its cool temperature, very high luminosity, variability properties, and spectroscopic properties, it would most likely be a sAGB star. Better constraints on the mass and Ca abundance of HV\,2112, as well as progress in theoretical modelling of T\.ZOs are needed to fully determine its true nature.

\acknowledgements

The authors thank: Emma Beasor, Jo Bovy, Ben Davies, Nancy Remage Evens, Emily Levesque, Chris Matzner, Phil Massey, Nidia Morrel, Krzysztof Stanek, Todd Thompson, and Dennis Zaritsky for helpful discussions; Dan Kelson, Jennifer Laing, Adiv Paradise, and Ayushi Singh for technical help; Tana Joseph \& Miroslav Filipovic for access to ASKAP-EMU images; Naomi McClure-Griffiths for access to SMC HI data. We thank the referee for helpful comments on the manuscript.

Spectroscopic data in this paper was obtained at the 6.5m Magellan Telescope at Las Campanas Observatory, Chile. The authors would like to thank all of the staff at LCO for their expertise and help.

The authors at the University of Toronto acknowledge that the land on which the University of Toronto operates is the traditional territory of the Huron-Wendat, the Seneca, and the Mississaugas of the Credit River. They are grateful to have the opportunity to work on this land.

The Dunlap Institute is funded through an endowment established by the David Dunlap family and the University of Toronto.

A.J.G.O. acknowledges support from the Lachlan Gilchrist Fellowship Fund. MRD acknowledges support from the NSERC through grant RGPIN-2019-06186, the Canada Research Chairs Program, the Canadian Institute for Advanced Research (CIFAR), and the Dunlap Institute at the University of Toronto. B.M.G. acknowledges the support of the Natural Sciences and Engineering Research Council of Canada (NSERC) through grant RGPIN-2022-03163, and of the Canada Research Chairs program. CSK is supported by NSF grants AST-1908570 and AST-1814440. B.J.S. is supported by NSF grants AST-1907570, AST-1908952, AST-1920392, and AST-1911074.

Support for this work was provided by NASA through the NASA Hubble Fellowship Program grant \#HST-HF2-51457.001-A awarded by the Space Telescope Science Institute, which is operated by the Association of Universities for Research in Astronomy, Inc., for NASA, under contract NAS5-26555.

This research has made use of the following software: \software{astropy \citep{Astropy.Collab.2018.Astropy}, IRAF \citep{Tody.D.1986.IRAFI,Tody.D.1993.IRAFII}, TOPCAT \citep{Taylor.M.2005.TOPCAT}, MARCS \citep{Gustafsson.B.2008.MARCSModels}}

\clearpage
\appendix

\restartappendixnumbering

\section{Kinematic Distribution of Total Proper Motion Relative to the Local Mean}\label{sec:totalpm}

\begin{figure*}[ht]
    \centering
    \includegraphics[width=0.75\linewidth]{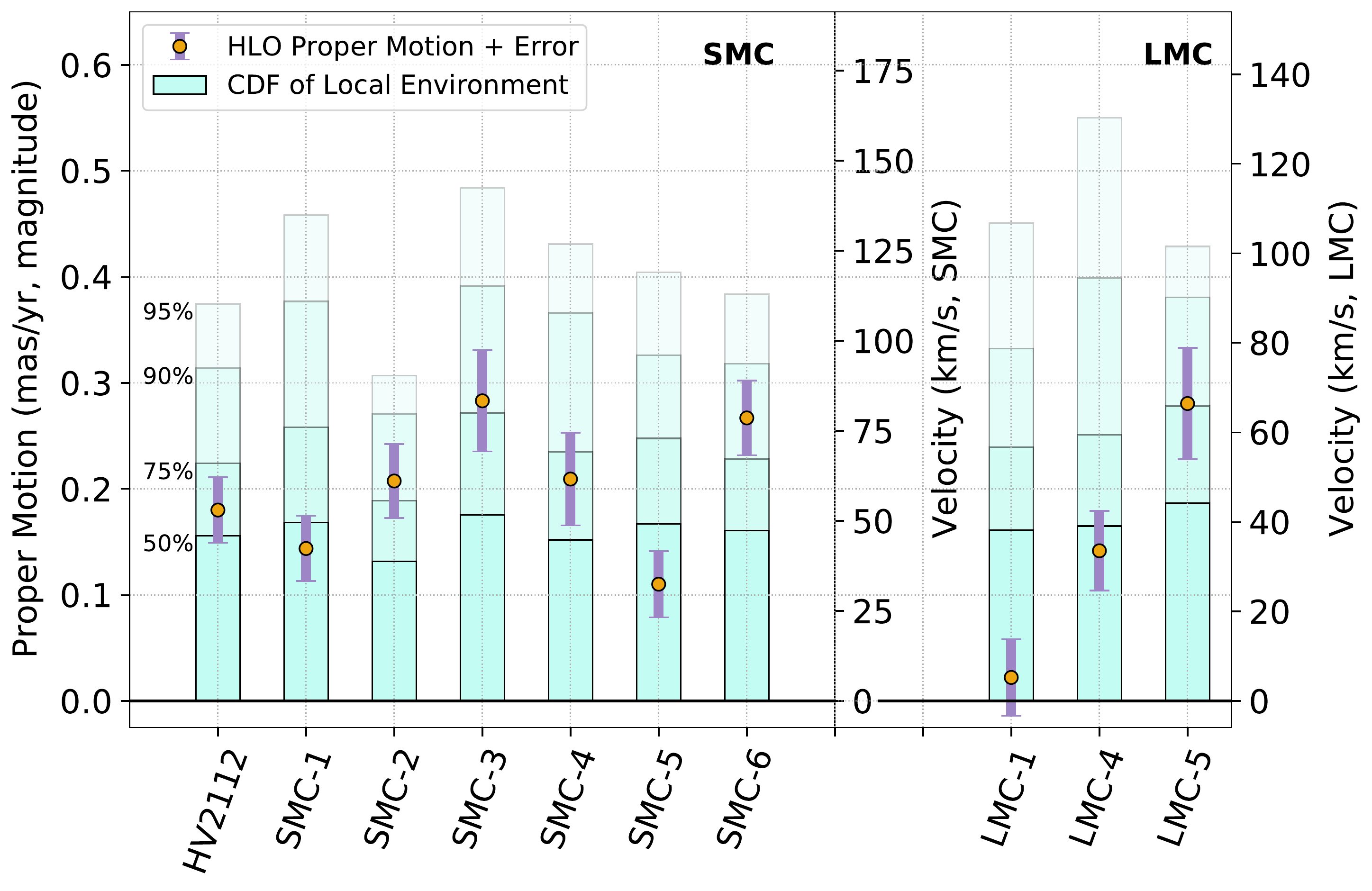}
    \caption{A comparison of the magnitude of the HLO proper motion to their local environments. The motion of the S/LMC and the weighted mean proper motion for the 5$'$ area around each HLO has been subtracted, removing bulk motion and allowing the proper motion of each HLO to be compared to the average dispersion of their local environment. The segmented light blue bars denote 50/75/90/95\% of the cumulative distribution function. The magnitude of the proper motion and associated measurement uncertainty are plotted for each HLO as a gold circle and violet error bar. Velocities (for SMC and LMC distances) are shown on the right axes.}
    \label{fig:sagb_whisker} 
\end{figure*}

We also assess the magnitude of the total proper motion of the HLOs/HAVs relative to their local environments. After subtracting the weighted means listed in Table~\ref{tab:kinematics}, all of the HLOs except LMC-1 have residual proper motions of 0.1--0.3 mas yr$^{-1}$. In Figure~\ref{fig:sagb_whisker} we show where the HLOs lie relative to the cumulative distribution of all the nearby stars. Four HLOs (40\%) fall within the inner 50\% of the 1-D CDF, while 4 (40\%) also fall in outer 25\%. For the HAVs, 12 (44\%) and 7 (26\%) fall within the inner 50\% and outer 25\% of their local CDFs, respectively.

In order to ascertain whether the fraction of HLOs/HAVs that fall in the outer regions of their local CDFs is actually higher than would expected given the quality of their astrometric solutions, we performed the same analysis with a sample of other stars throughout the Magellanic Clouds. In each galaxy we randomly chose 10 test stars with two constraints: the star had to (i) be a probable Cloud member, and (i) have astrometric excess noise in the range $0.2 < \epsilon_{i} < 0.5$ with a significance $D > 2$ (the same as the sample of HLOs). We then queried \emph{Gaia} EDR3 for all stars within 5$'$ of each test star, removed foreground contamination, and determined the location of the star within its local total proper motion CDF.

We ran this trial 500 times in each Cloud and found that in $\sim$54\% of these trials, 4/10 sources lie in the outer 25\% of the 1-D CDF. This is similar to the results of the 10 HLOs. Thus, we do not see evidence that the HLOs have unusual total proper motions relative to their local mean.

\clearpage

\section{SFH Cepheid Test}\label{sfhceph}

\begin{figure*}[h!]
    \centering
    \includegraphics[width=0.95\textwidth]{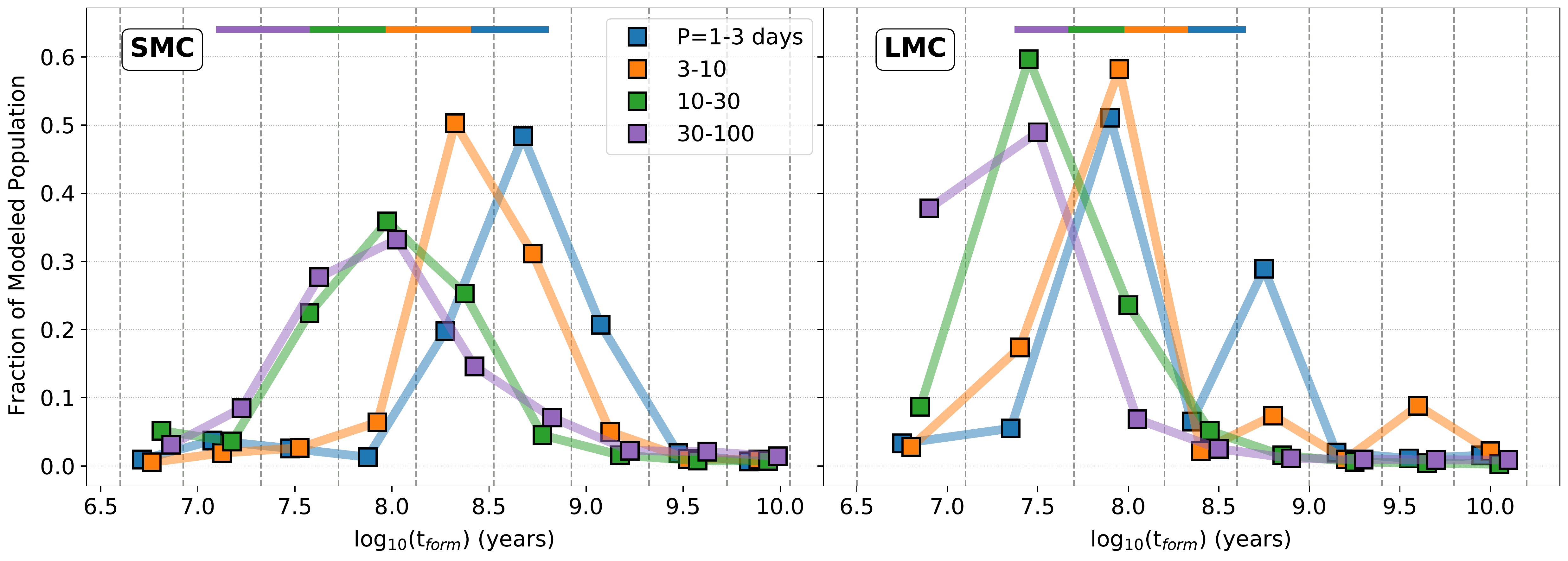}
    \caption{The same as Figure \ref{fig:sfh_hlosRSGsAGBs} but plotting a population of OGLE-III Cepheid variables. The populations are organized by variability period: 1-3 days in blue, 3-10 days in orange, 10-30 days in green, and 30-100 days in violet. The expected age range for each set of periods, calculated from \citet{Anderson.R.2016.CepheidPeriodAge}, is shown as a colored bar at the top of the plot.}
    \label{fig:sfh_cepheids}
\end{figure*}

\section{SFH Quantitative Likelihoods}\label{sfhtable}

Here we quantify the similarity in star formation history of stellar populations by computing three likelihoods.  We first find the maximum likelihood models for population $1$ and $2$, $L_1$ and $L_2$.  Then we find the maximum likelihood model $L_{12}=L_1' L_2'$ for modeling them simultaneously assuming they have the same association with stellar ages by computing their likelihoods $L_1'$ and $L_2'$ for the same $\epsilon_i$ and maximizing $L_{12}$. We renormalize the $\epsilon_i$ for the two populations separately since they will generally contain different numbers of objects. The logarithmic difference, $\Delta\ln L = \ln L_{12} - \ln L_1 - \ln L_2$ is then a quantitative estimate of their similarity, yielding $\Delta \ln L = 0$ if there age distributions are identical. For test problems with $50$ stars randomly distributed following the star formation rate at a fixed age, this approach robustly identifies random samples with the same age even though the random realizations of the populations are different. In Table \ref{sfh:tab} we present the likelihoods for all populations shown in Figure \ref{fig:sfh_hlosRSGsAGBs}. The populations with the smallest $\Delta \ln L$ values for the HLOs and HAVs are AGB stars and not RSGs or associated high-mass phenomena (SNRs/HMXBs). This is consistent with our analysis in \S\ref{sec:sfh}.

\begin{deluxetable}{r|ccc|ccc}[ht!]\label{sfh:tab}
\centering
\tablecaption{Likelihood of similarity for observed populations}
\tablehead{\multicolumn{1}{r|}{Sample} & \multicolumn{3}{c|}{HLO} & \multicolumn{3}{c}{HLO+HAV}  \\
  \multicolumn{1}{r|}{}  & \colhead{LMC} & \colhead{SMC} & \multicolumn{1}{c|}{Tot} & \colhead{LMC} & \colhead{SMC} & \multicolumn{1}{c}{Tot}
}

\startdata
HLO               & -0.202 &  0.001 & -0.202 &  1.264 & -0.011 &  1.253  \\
HLO/HAV           &  1.382 & -0.054 &  1.328 & -0.182 & -0.124 & -0.306  \\\hline
O Rich AGB        &  2.672 &  0.403 &  3.075 &  2.607 &  0.372 &  2.978 \\
low L O Rich AGB  &  2.633 &  0.020 &  2.654 &  2.854 &  0.173 &  3.027  \\
high L O Rich AGB &  2.731 &  0.150 &  2.881 &  4.336 &  1.224 &  5.560  \\\hline
RSG               &  5.976 &  1.959 &  7.934 & 20.162 &  3.890 & 24.052  \\
SNR               &  2.942 &  3.554 &  6.496 &  5.396 &  6.071 & 11.467  \\
HMXB              &  2.477 &  4.777 &  7.254 &  4.356 &  6.180 & 10.536  \\\hline\hline
\enddata
\end{deluxetable}

\clearpage
\section{Radio Surface Brightness Upper Limits}\label{hav_radio}

\begin{deluxetable}{ccc|cccccc}[ht!]
\tabletypesize{\footnotesize}
\centering
\tablecaption{Radio Surface Brightness Upper Limits\label{tab:radio_ul_all}}
\tablehead{\colhead{Name (HLO)} & \colhead{RA} & \multicolumn{1}{c|}{Dec} & \colhead{ASKAP} & \colhead{ASKAP} & \colhead{ASKAP} & \colhead{ATCA} & \colhead{ATCA} & \colhead{ATCA}\vspace{-0.25cm}\\ 
\colhead{} & \colhead{(deg)} & \multicolumn{1}{c|}{(deg)} & \colhead{887 MHz} & \colhead{960 MHz} & \colhead{1320 MHz} & \colhead{2734 MHz} & \colhead{4800 MHz} & \colhead{8600 MHz}
}
\startdata
\multicolumn{8}{c}{SMC}\\\hline
HV 2112 & 17.515856 & $-$72.614603 & & 0.2 & 0.4 & 0.6 & 0.8 & 4.5 \\ 
SMC-1 & 11.703220 & $-$72.763824 & & 0.2 & 0.6 & 0.5 & 0.9 & 4.0 \\ 
SMC-2 & 13.036803 & $-$71.606606 & & 0.2 & 0.3 & 0.5 & 0.8 & 4.7 \\ 
SMC-3 & 13.909812 & $-$73.194845 & & 0.2 & 0.4 & 0.5 & 0.7 & 4.4 \\
SMC-4 & 15.403412 & $-$72.744751 & & 0.2 & 0.5 & 0.8 & 0.9 & 3.9 \\
SMC-5 & 15.903689 & $-$73.560525 & & 0.2 & 0.4 & 0.6 & 0.8 & 4.8 \\
SMC-6 & 17.612562 & $-$72.596670 & & 0.3 & 0.4 & 0.5 & 0.8 & 4.6 \\\hline
& 10.339302 & $-$72.837669 & & 0.2 & 0.4 & 0.5 & 0.8 & 4.6 \\
& 14.310338 & $-$73.022591 & & 0.6 & 1.2 & 0.9 & 0.9 & 4.8 \\
& 14.709024 & $-$72.309883 & & 0.5 & 0.6 & 3.3 & 0.9 & 3.7 \\\hline
\multicolumn{8}{c}{LMC}\\\hline
LMC-1 & 80.824095 & $-$66.952095 & 0.6 & & & & 0.8 & 4.4  \\
LMC-3 & 84.986223 & $-$69.589172 & 8.0 & & & & 2.5 & 8.3 \\
LMC-4 & 86.709478 & $-$67.246312 & 0.4 & & & & 1.3 & 4.3 \\
LMC-5 & 88.116079 & $-$69.236122 & 0.8 & & & & 0.9 & 4.3\\\hline
& 68.98892 & -70.997383 & 0.3 & & & & X & X \\
& 70.108388 & -71.661774 & 0.4 & & & & X & X \\
& 73.436919 & -68.966476 & 0.5 & & & & 1.2 & 4.5 \\
& 74.370204 & -70.458191 & 0.4 & & & & 1.0 & 4.1 \\
& 74.731888 & -66.76152 & 0.7 & & & & 0.7 & 4.1 \\
& 75.349969 & -70.098244 & 0.6 & & & & 0.9 & 4.5 \\
& 76.517648 & -70.28093 & 1.0 & & & & 0.8 & 4.5 \\
& 76.615336 & -68.201019 & 1.5 & & & & 1.7 & 4.3 \\
& 76.664509 & -71.599014 & 0.4 & & & & 0.8 & 3.9 \\
& 76.909604 & -69.735832 & 1.0 & & & & 0.9 & 4.0 \\
& 77.999464 & -71.606895 & 0.3 & & & & 0.7 & 4.5 \\
& 78.92018 & -66.08271 & 0.6 & & & & 1.0 & 4.5 \\
& 79.793567 & -70.972542 & 0.3 & & & & 0.8 & 4.6 \\
& 80.006549 & -67.578377 & 1.1 & & & & 0.8 & 3.7 \\
& 80.79228 & -67.835068 & 2.6 & & & & 1.2 & 4.3 \\
& 81.092492 & -66.110359 & 0.9 & & & & 0.8 & 4.0 \\
& 81.138056 & -70.71003 & 0.4 & & & & 0.8 & 4.1 \\
& 82.323732 & -67.042938 & 0.9 & & & & 0.9 & 4.5 \\
& 83.249667 & -70.68988 & 4.7 & & & & 7.7 & 21.8 \\
& 85.173783 & -66.246323 & 0.5 & & & & 0.7 & 4.2 \\
& 87.305663 & -70.711296 & 0.5 & & & & 0.5 & 3.3 \\
& 87.980191 & -71.078644 & 0.4 & & & & 0.6 & 3.3 \\
& 89.684721 & -68.447166 & 0.5 & & & & X & X \\
& 91.290672 & -72.676445 & 0.4 & & & & X & X \\\hline\hline
\enddata
\tablecomments{Units: $\Sigma_{1 GHz}$ 1e-21 W m$^{-2}$ Hz$^{-1}$ sr$^{-1}$ \\ X denotes star outside of map boundaries \\ A blank space denotes the star not being in that map}
\end{deluxetable}

\clearpage

\section{Temperature and pseudo-Equivalent Width Values}\label{sec:pew_t_final}

\begin{deluxetable}{c|cccccccccc}[ht!]\label{tab:pew_t_final}

\centering
\tablecaption{Temperatures and pseudo-equivalent width values\label{tab:tempeqw}}
\tablehead{\multicolumn{1}{c|}{Name} & \colhead{T$_{\mathrm{eff}}$ (K)} & \colhead{Fe 5569\AA} & \colhead{Mo 5570\AA} & \colhead{Ca 6572\AA} & \colhead{Li 6707\AA} & \colhead{K 7698\AA} & \colhead{Ni 7797\AA} & \colhead{Rb 7800\AA} & \colhead{Fe 7802\AA}
}
\startdata
\multicolumn{9}{c}{HLOs}\\\hline
HV 2112 & 3588$^{70}_{110}$ & 0.13$\pm$0.01 & 0.09$\pm$0.0 & 0.28$\pm$0.01 & 0.32$\pm$0.0 & 0.21$\pm$0.0 & 0.11$\pm$0.01 & 0.11$\pm$0.0 & 0.06$\pm$0.0 \\
LMC-1 & 3473$^{110}_{90}$ & 0.21$\pm$0.01 & 0.02$\pm$0.01 & 0.27$\pm$0.01 & 0.35$\pm$0.01 & 0.17$\pm$0.02 & 0.07$\pm$0.0 & 0.09$\pm$0.01 & 0.1$\pm$0.01 \\
LMC-3 & 3462$^{80}_{70}$ & 0.24$\pm$0.01 & 0.07$\pm$0.0 & 0.24$\pm$0.0 & 0.07$\pm$0.0 & 0.15$\pm$0.0 & 0.09$\pm$0.0 & 0.1$\pm$0.01 & 0.1$\pm$0.01 \\
LMC-4 & 3456$^{80}_{60}$ & 0.21$\pm$0.01 & 0.02$\pm$0.01 & 0.23$\pm$0.0 & 0.22$\pm$0.0 & 0.19$\pm$0.0 & 0.11$\pm$0.01 & 0.16$\pm$0.03 & 0.16$\pm$0.02 \\
SMC-1 & 3621$^{40}_{110}$ & X & X & 0.23$\pm$0.01 & 0.34$\pm$0.0 & 0.28$\pm$0.0 & 0.07$\pm$0.0 & 0.09$\pm$0.01 & 0.07$\pm$0.01 \\
SMC-2 & 3517$^{80}_{80}$ & 0.19$\pm$0.01 & 0.03$\pm$0.0 & 0.21$\pm$0.01 & 0.47$\pm$0.0 & 0.25$\pm$0.0 & 0.04$\pm$0.0 & 0.05$\pm$0.0 & 0.04$\pm$0.0 \\
SMC-3 & 3541$^{110}_{110}$ & 0.27$\pm$0.01 & 0.1$\pm$0.0 & 0.32$\pm$0.0 & 0.58$\pm$0.0 & 0.33$\pm$0.0 & 0.1$\pm$0.0 & 0.05$\pm$0.0 & 0.04$\pm$0.0 \\
SMC-5 & 3462$^{110}_{80}$ & 0.19$\pm$0.01 & 0.09$\pm$0.01 & 0.29$\pm$0.01 & 0.54$\pm$0.01 & 0.3$\pm$0.0 & 0.07$\pm$0.01 & 0.11$\pm$0.01 & X \\
SMC-6 & 3439$^{80}_{50}$ & 0.11$\pm$0.02 & 0.02$\pm$0.01 & 0.08$\pm$0.0 & 0.04$\pm$0.0 & 0.08$\pm$0.0 & 0.06$\pm$0.0 & 0.12$\pm$0.01 & 0.07$\pm$0.0 \\\hline
\multicolumn{9}{c}{HAVs}\\\hline
HAV-1 & 3388$^{80}_{50}$ & 0.31$\pm$0.01 & 0.04$\pm$0.0 & 0.21$\pm$0.01 & 0.34$\pm$0.0 & 0.18$\pm$0.0 & 0.11$\pm$0.01 & 0.23$\pm$0.01 & 0.13$\pm$0.01 \\
HAV-2 & 3338$^{80}_{0}$ & 0.23$\pm$0.02 & 0.03$\pm$0.01 & 0.23$\pm$0.02 & 0.31$\pm$0.01 & 0.17$\pm$0.0 & 0.14$\pm$0.0 & 0.19$\pm$0.01 & 0.17$\pm$0.01 \\
HAV-3 & 3498$^{30}_{80}$ & 0.22$\pm$0.02 & 0.03$\pm$0.01 & 0.21$\pm$0.01 & 0.24$\pm$0.01 & 0.27$\pm$0.0 & 0.09$\pm$0.0 & 0.12$\pm$0.01 & 0.11$\pm$0.01 \\
HAV-4 & 3497$^{30}_{80}$ & 0.12$\pm$0.02 & 0.02$\pm$0.01 & X & 0.1$\pm$0.01 & 0.05$\pm$0.0 & 0.07$\pm$0.01 & 0.09$\pm$0.02 & 0.1$\pm$0.02 \\\hline
\multicolumn{9}{c}{RSGs}\\\hline
RSG-1 & 3388 & 0.174$\pm$0.005 & 0.039$\pm$0.003 & 0.291$\pm$0.003 & 0.03$\pm$0.003 & 0.458$\pm$0.001 & 0.108$\pm$0.001 & 0.064$\pm$0.002 & 0.075$\pm$0.007 \\
RSG-2 & 3908 & 0.331$\pm$0.002 & 0.106$\pm$0.004 & 0.388$\pm$0.004 & 0.018$\pm$0.005 & 0.435$\pm$0.001 & 0.164$\pm$0.001 & 0.042$\pm$0.003 & 0.047$\pm$0.003 \\
RSG-3 & 3767 & 0.363$\pm$0.002 & 0.1$\pm$0.003 & 0.387$\pm$0.004 & 0.044$\pm$0.003 & 0.405$\pm$0.003 & 0.17$\pm$0.002 & 0.016$\pm$0.001 & 0.023$\pm$0.001 \\
RSG-4 & 3664 & 0.254$\pm$0.01 & 0.091$\pm$0.005 & 0.29$\pm$0.004 & 0.03$\pm$0.004 & 0.451$\pm$0.001 & 0.17$\pm$0.004 & 0.037$\pm$0.003 & 0.056$\pm$0.003 \\
RSG-5 & 3572 & 0.235$\pm$0.012 & 0.091$\pm$0.007 & 0.272$\pm$0.003 & 0.031$\pm$0.006 & 0.436$\pm$0.002 & 0.151$\pm$0.004 & 0.05$\pm$0.004 & 0.061$\pm$0.004 \\
RSG-6 & 3564 & 0.147$\pm$0.008 & 0.055$\pm$0.004 & 0.262$\pm$0.004 & 0.041$\pm$0.008 & 0.515$\pm$0.002 & 0.105$\pm$0.001 & 0.055$\pm$0.003 & 0.067$\pm$0.005 \\
RSG-7 & 3655 & 0.378$\pm$0.004 & 0.104$\pm$0.002 & 0.38$\pm$0.002 & 0.119$\pm$0.001 & 0.332$\pm$0.001 & 0.169$\pm$0.0 & 0.027$\pm$0.001 & 0.033$\pm$0.001 \\
RSG-8 & 3630 & 0.332$\pm$0.009 & 0.107$\pm$0.005 & 0.343$\pm$0.003 & 0.021$\pm$0.002 & 0.363$\pm$0.004 & 0.199$\pm$0.013 & 0.038$\pm$0.007 & 0.049$\pm$0.007 \\
RSG-9 & 3706 & 0.219$\pm$0.011 & 0.085$\pm$0.008 & 0.293$\pm$0.003 & 0.056$\pm$0.008 & 0.373$\pm$0.005 & 0.158$\pm$0.007 & 0.049$\pm$0.005 & 0.059$\pm$0.003 \\
RSG-10 & 3935 & 0.342$\pm$0.006 & 0.103$\pm$0.003 & 0.337$\pm$0.003 & 0.037$\pm$0.002 & 0.332$\pm$0.004 & 0.167$\pm$0.004 & 0.024$\pm$0.001 & 0.036$\pm$0.003 \\
RSG-11 & 3491 & 0.327$\pm$0.011 & 0.088$\pm$0.008 & 0.344$\pm$0.003 & 0.052$\pm$0.005 & 0.48$\pm$0.002 & 0.165$\pm$0.004 & 0.015$\pm$0.003 & 0.045$\pm$0.003 \\
RSG-12 & 3539 & 0.352$\pm$0.011 & 0.141$\pm$0.008 & 0.352$\pm$0.004 & 0.047$\pm$0.004 & 0.421$\pm$0.004 & 0.157$\pm$0.005 & 0.018$\pm$0.004 & 0.033$\pm$0.003 \\
RSG-13 & 3427 & 0.285$\pm$0.004 & 0.088$\pm$0.003 & 0.396$\pm$0.007 & 0.018$\pm$0.004 & 0.365$\pm$0.002 & 0.178$\pm$0.017 & 0.034$\pm$0.005 & 0.04$\pm$0.009 \\
RSG-14 & 3483 & 0.35$\pm$0.005 & 0.112$\pm$0.003 & 0.399$\pm$0.003 & 0.094$\pm$0.003 & 0.404$\pm$0.002 & 0.164$\pm$0.003 & 0.016$\pm$0.006 & 0.029$\pm$0.003 \\
\enddata
\tablecomments{RSG temperatures from \citep{Neugent.K.2012.RSG.YSG.LMC}. Temperature and pseudo-equivalent width measurements described in \S\ref{spec_pew_measure}.}
\end{deluxetable}

\section{Notes on Specific Line Measurements}\label{sec:spec_line_note}

\emph{K I}: The K I 7698.97$\mathrm{\text{\AA}}$ feature regularly appeared as a doublet in our observations. This doublet is thought to contain both stellar and interstellar components \citep{Herbig.G.1982.KInterstellarLine}. The stellar component would have an RV similar to that LMC compared to the interstellar component. We take only the bluest component of the feature for all of our measurements. 

\emph{Li I}: Li I 6707.91$\mathrm{\text{\AA}}$ often displays a blended Fe I 6707.43$\mathrm{\text{\AA}}$ component (most obviously in HV\,2112). We are careful to separate this Fe I line when possible and do not include it in the Li I measurement. Both NIST ASD and AtLL identify two lines of Li I in close proximity: 6707.76$\mathrm{\text{\AA}}$ and 6707.91$\mathrm{\text{\AA}}$. We allow $\delta\lambda$ between the Li I 6707.91$\mathrm{\text{\AA}}$ and Fe I 6707.43$\mathrm{\text{\AA}}$ features to vary over 0.33--0.48$\mathrm{\text{\AA}}$. 

\emph{Rb I}: For Rb I 7800.23$\mathrm{\text{\AA}}$, AtLL reports a nearby line of Fe I at 7799.535$\mathrm{\text{\AA}}$. This partially blended line is clearly visible in the spectrum of HV\,2112, and also appears throughout the rest of our data. We take care to separate this from the Rb I measurement. Finally, we found a shift in this Fe I--Rb I blend relative to the nearby Ni I 7797.58$\mathrm{\text{\AA}}$ and Fe I 7802.47$\mathrm{\text{\AA}}$ lines. When comparing our spectrum of HV\,2112 to Figure 2 panel (e) of \citet{Levesque.E.2014.HV2112disc}, the same pattern of lines is easily identifiable. However, throughout \textit{all} of our spectra, the measured location of the Fe I--Rb I blend is slightly redshifted when compared to the measured locations of the adjacent Fe I and Ni I lines (i.e. for Ni I 7797.58 and Rb I 7800.23, $\Delta\lambda >$ 2.65$\text{\AA}$). For the HLOs and HAVs, Rb I is consistently $\sim$0.2--0.3$\mathrm{\text{\AA}}$ redder  (a velocity change of $\sim$8--12 km/s). For RSGs it is more pronounced, with a consistent shift of $\sim$0.55-0.65$\mathrm{\text{\AA}}$ redder ($\delta$v$\sim$21--25km/s). This may mean that the Fe I-Rb I absorption is from a different level of the photosphere than the adjacent Ni I and Fe I lines.

\bibliography{thesis_bib}
\bibliographystyle{aasjournal}

\end{CJK*}
\end{document}